\documentclass[aps,pra,superscriptaddress,reprint]{revtex4-2}
\usepackage{amsmath}
\usepackage{amssymb}
\usepackage{mathrsfs}
\usepackage{graphicx}
\usepackage{booktabs}  
\usepackage{multirow}  
\usepackage[colorlinks=true, letterpaper=true, pdfstartview=FitV, linkcolor=blue, citecolor=blue, urlcolor=blue]{hyperref}

\begin{document}
	
	\title{Dynamical Blockade Optimizing via Particle Swarm Optimization Algorithm}

	\author{Guang-Yu Zhang}
	\affiliation{Key Laboratory of Low-Dimensional Quantum Structures and
		Quantum Control of Ministry of Education, Key Laboratory for Matter
		Microstructure and Function of Hunan Province, Department of Physics and
		Synergetic Innovation Center for Quantum Effects and Applications, Hunan
		Normal University, Changsha 410081, China}

\author{Zhi-Hao Liu}
\affiliation{Key Laboratory of Low-Dimensional Quantum Structures and
Quantum Control of Ministry of Education, Key Laboratory for Matter
Microstructure and Function of Hunan Province, Department of Physics and
Synergetic Innovation Center for Quantum Effects and Applications, Hunan
Normal University, Changsha 410081, China}

\author{Xun-Wei Xu}
\email{xwxu@hunnu.edu.cn}
\affiliation{Key Laboratory of Low-Dimensional Quantum Structures and
Quantum Control of Ministry of Education, Key Laboratory for Matter
Microstructure and Function of Hunan Province, Department of Physics and
Synergetic Innovation Center for Quantum Effects and Applications, Hunan
Normal University, Changsha 410081, China}
\affiliation{Institute of Interdisciplinary Studies, Hunan Normal University, Changsha, 410081, China}
	
	\date{\today}
	
	\begin{abstract}
Photon blockade in weak nonlinear regime is an exciting and promising subject that has been extensively studied in the steady state.
However, how to achieve dynamic blockade in a \emph{single} bosonic mode with \emph{weak} nonlinearity using \emph{only} pulsed driving field remains unexplored.
Here, we propose to optimize the parameters of the pulsed driving field to achieve dynamic blockade in a single bosonic mode with weak nonlinearity via the particle swarm optimization (PSO) algorithm.
We demonstrate that both Gaussian and rectangular pulses can be used to generate dynamic photon blockade in a single bosonic mode with weak nonlinearity.
Based on the Fourier series expansions of the pulsed driving field, we identify that there are many paths for two-photon excitation in the bosonic mode, even only driven by pulsed field, and the dynamic blockade in weak nonlinear regime is induced by the destructive interference between them.
Our work not only highlights the effectiveness of PSO algorithm in optimizing dynamical blockade, but also opens a way to optimize the parameters for other quantum effects, such as quantum entanglement and quantum squeezing.
	\end{abstract}
	
	\maketitle

	\section{Introduction}
	
	Photon blockade is a pure quantum effect that suppresses multiphoton generation in a bosonic mode~\cite{PhysRevLett.79.1467}. It is a mechanism for the generation of single photons through coherent optical driving~\cite{Buckley_2012,RevModPhys.87.347},	and plays a pivotal role in the development of quantum computing~\cite{RN565,RevModPhys.79.135}, quantum networks~\cite{RN566,RevModPhys.87.1379}, quantum cryptography~\cite{RevModPhys.81.1301,RevModPhys.92.025002}
and quantum sensing~\cite{RevModPhys.89.035002,Pirandola2018NaPho}. 
Conventionally, photon blockade is proposed by introducing various strong nonlinear interactions into the optical modes~\cite{Ridolfo2012PRL,LiuYX2014PRA,Adam2014PRA,PhysRevLett.107.063601,PhysRevLett.107.063602,LiaoJQ2013PRA,XieH2016PRA,Majumdar2013PRB,Huang2018PRL,Huang2022LPRv,Chakram2022NatPh,Zhou2020PRA}, and has been observed in some experimental platforms, such as optical
cavities coupled to single atoms~\cite{Birnbaum2005Natur,Dayan2008Sci,Aoki2009PRL}, quantum dots embedded in photonic-crystal nanocavities~\cite{Faraon2008NatPh,Reinhard2012NaPho}, and superconducting qubits resonantly coupled to microwave resonators~\cite{LangC2011PRL,Hoffman2011PRL}.
However, strong nonlinearity is still difficult to achieve in most experimental platforms, and photon blockade in weak nonlinear regime is an exciting and promising subject.

In the past decade, some novel mechanisms are proposed to obtain photon blockade under weak nonlinear interactions. 
The one that attracts the most attention is the photon blockade originated from the suppression of two-photon excitation by destructive interference between different transition paths. It was firstly predicted in weakly nonlinear photonic molecules~\cite{PhysRevLett.104.183601,PhysRevA.83.021802,PhysRevA.83.021802} and then observed for both optical~\cite{Snijders2018PRL} and microwave~\cite{Vaneph2018PRL} photons. Photon blockade based on destructive interference has been extensively studied in various systems, such as coupled optomechanical systems~\cite{Xu_2013,savona2013UPB,ZhangWZ2015PRA}, coupled cavities with second or third order nonlinearities~\cite{Ferretti_2013,PhysRevA.88.033836,Gerace2014PRA,PhysRevA.90.043822,PhysRevA.90.033809,Lemonde2014PRA,PhysRevA.91.063808,ZhouYH2015PRA,PhysRevA.96.053810,PhysRevA.96.053827,PhysRevA.104.053718,ShenS2019PRA,Zubizarreta2020LPR,Wang_2020NJP}, cavity embedded with a quantum dot~\cite{PhysRevLett.108.183601,ZhangW2014PRA,TangJ2015NatSR,LiangXY2019PRA,PhysRevLett.125.197402}, coupled-resonator chain~\cite{WangY2021PRL,li2023enhancement,lu2024chiral}, etc.
Besides, photon blockade under weak nonlinear interactions has also been predicted in a bosonic mode with nonlinear driving~\cite{Andrew2021sciadv,MaYX2023PRA} or nonlinear loss~\cite{ZhouYH2022PRAPP,SuX2022PRA,Ben-Asher2023PRL,ZuoYL2022PRA}.
	

Different from the photon blockade predicted in the steady state by constant driving, dynamical blockade has been proposed for the nonlinear systems driven by a field with time-dependent amplitude~\cite{PhysRevA.93.043857,PhysRevA.90.013839,PhysRevA.90.063805,PhysRevB.97.241301,PhysRevLett.123.013602,PhysRevLett.129.043601} or using time-dependent coupling~\cite{Stefanatos2020PRA}.	
Due to the destructive interference between different paths for two-photon excitation, dynamical blockade has been predicted in the weakly nonlinear regime when a bosonic mode is coupled to two other modes by four-wave mixing~\cite{PhysRevA.90.063805}, or coupled to a gain medium~\cite{PhysRevB.97.241301}, or driven by a combination of both continuous and pulsed fields~\cite{PhysRevLett.123.013602}, or driven by a bi-tone coherent field~\cite{PhysRevLett.129.043601}.
Nevertheless, how to achieve dynamic blockade in a single bosonic mode with weak nonlinearity driven by \emph{only} pulsed field is still an open question.

In this paper, we propose to optimize the parameters of the pulsed driving field to achieve dynamic blockade in a single bosonic mode with weak nonlinearity via the particle swarm optimization (PSO) algorithm~\cite{488968}.
PSO algorithm, inspired by collective behaviors observed in natural phenomena such as flocks of birds and schools of fish, operates as a group stochastic optimization algorithm, which has expensive applications in physics, such as crystal structure prediction~\cite{Wang2010PRB}, design of diffraction grating filters~\cite{Shokooh-Saremi:07}, maximization of topological invariants~\cite{PhysRevB.100.235452}, characterization of dephasing quantum systems~\cite{PhysRevA.93.012122}, and cosmological parameter estimation~\cite{Prasad2012PRD}.
Here, we apply the PSO algorithm to demonstrate dynamic blockade in a single bosonic mode with weak Kerr nonlinearity using \emph{only} pulsed excitations.
We optimize the parameters of the Gaussian and rectangular pulses by the PSO algorithm to generate strong dynamic blockade in a single bosonic mode with weak nonlinearity.
The method of parameters optimization based on PSO algorithm is universal, which can also be applied to optimize the parameters for quantum entanglement and quantum squeezing~\cite{LiuYH2024arxiv}.

	The paper is organized as follows. 
	In Sec.~\ref{BIN}, we present the model of a single bosonic mode with Kerr nonlinearity driven by a pulsed coherent field, and briefly introduce the PSO algorithm. 
	We demonstrate the dynamic blockade in a bosonic mode driven by a series of Gaussian pulses optimized by the PSO algorithm in Sec.~\ref{CIN}.
	In Sec.~\ref{WBON}, we show that dynamic blockade also can be observed in the bosonic mode when it is driven by a series of optimized rectangular pulses. 
	Finally, we make the conclusions in Sec.~\ref{Con}.

	\section{Model and algorithm}\label{BIN}
	\subsection{Physical model}
	We consider a single bosonic mode with Kerr nonlinearity driven by a pulsed coherent field, and in the frame rotating with the driving frequency $\omega_d$, the system can be described by the Hamiltonian ($\hbar=1$)	
	\begin{equation}\label{eq1}
		H=\Delta a^{\dagger}a+Ua^{\dagger}a^{\dagger}aa+\varepsilon (t)(a^{\dagger}+a),
	\end{equation}
	where $a^{\dagger}(a)$ is the creation (annihilation) operator of the bosonic mode with the resonant frequency $\omega_a$, $U$ is the strength of nonlinearity, 
$\varepsilon (t)$ is the envelope of the pulsed driving field at time $t$, and $\Delta\equiv \omega_a- \omega_d $ is the detuning between the bosonic mode and driving field.
	
The dynamic of the system is governed by the quantum master equation~\cite{Carmichael1993}
\begin{equation}\label{MasterEq}
	\dot{\rho }=-i \left[ H,\rho%
	\right]  +\frac{\gamma }{2}(2a\rho 
	a^{\dag }-a^{\dag }a\rho  -\rho  a^{\dag }a ),
\end{equation}
where $\rho$ is the density matrix of the system and $\gamma$ is the decay rate of the bosonic mode. 
The statistics of the bosonic mode can be evaluated by the instantaneous equal-time second-order correlation function
\begin{equation}
	g^{(2)}\left( t\right) =
	\frac{\left\langle a^{\dag}\left( t\right)a^{\dag}\left( t\right)a\left( t\right)a\left( t\right) \right\rangle}
	{\left\langle a^{\dag}\left( t\right)a\left( t\right) \right\rangle \left\langle a^{\dag}\left( t\right)a\left( t\right) \right\rangle},
\end{equation}
which indicates dynamic blockade by $g^{(2)}\left( t\right)<1$.

As the Hamiltonian is time-dependent, except a few cases (such as using a simple bi-tone drive~\cite{PhysRevLett.129.043601}), it is hard to analytically obtain the optimal conditions for the dynamic blockade.
Here, we try to overcome this challenge via an optimization algorithm widely used in artificial intelligence (see next subsection), and numerically demonstrate dynamic blockade in a single bosonic mode in the weak nonlinear regime, i.e., $U<\gamma$, driven \emph{only} by pulsed coherent field. Without loss of generality, we set $U/\gamma=0.05$ in the following numerical simulations.
	
	\subsection{Optimization algorithm}
	
In this subsection, we briefly survey the optimization algorithm, i.e., PSO algorithm, that we will use to optimize
the parameters to achieve dynamic blockade in a single bosonic mode in the weak nonlinear regime.
To be more specific, we will search for the optimal parameters of the pulsed driving field by PSO algorithm to achieve minimal second-order correlation function $g^{(2)}\left( t\right)$.

	
	In the PSO algorithm, there are $N_p$ computational agents, being referred to as particles, without any relation to the physical particles. Each particle is assigned a position and a velocity vector.
	Suppose that the position of the $i$th particle at step $k$ ($k=0,1,2,\cdots,N_k$) is:
	\begin{equation}
		X_i^{k}=[X_{i1}^{k},X_{i2}^{k},\cdots ,X_{id}^{k}]^T,
	\end{equation}
	and the velocity is:
	\begin{equation}
		V_i^{k}=[V_{i1}^{k},V_{i2}^{k},\cdots ,V_{id}^{k}]^T,
	\end{equation}
	where `$d$' denotes the dimensionality corresponding to the number of unknown parameters in the problem to be solved, and $N_k$ is the maximum number of iterations.
The quality of the position is characterized by a fitness value evaluated by an optimization function (fitness function) based on the predetermined criteria.
	The position and velocity of the $i$th particle are updated by the following way:
	\begin{equation}  
		X_{i}^{k+1} = X_{i}^{k} + V_{i}^{k+1},
	\end{equation}	
	\begin{equation}
		V_{i}^{k+1} = w  V_{i}^{k} + f_1  r_1  (P^k_{i} - X_{i}^{k}) + f_2  r_2  (G^k - X_{i}^{k}).
	\end{equation}  
	Here, $w$ represents the inertial weight, 
	exerting a significant influence on the algorithm's search capability. 
	A larger $w$ promotes global exploration, 
	while a smaller $w$ is conducive to local exploration.
	The parameters $f_1$ and $f_2$ correspond to the individual cognitive learning factor and social learning factor, respectively. Without loss of generality, we set $w=0.5$ and $f_1=f_2=1.5$ for numerical simulation.
	The random numbers $r_1$ and $r_2$, sampled from the interval $[0,1]$, introduce an element of randomness to enhance the search process.
	$P^{k}_{i}$ is the historical best position of the $i$th particle within $k$ steps for the fitness function achieving maximum/minimal value, and $G^k$ denotes the historical best position for the entire swarm within $k$ steps. 

	In this work, the fitness function is defined as the minimum of the instantaneous equal-time second-order correlation function, denoted by $g^{(2)}_{\rm min}\left( t\right)$, within a specified time range. The instantaneous equal-time second-order correlation function is calculated by numerically solving the master equation~(\ref{MasterEq}) with the open-source software QuTiP~\cite{QuTiP1,QuTiP2}, for the system driven by a pulsed field characterized by $d$-dimensional parameters. The set of $d$-dimensional parameters serves as the position vector of the particle, and after each iteration, a new set of parameters is generated. The stabilization of the particle's fitness value, i.e., $g^{(2)}_{\rm min}\left( t\right)$, indicates the attainment of an optimal position, with the corresponding best position $G^k$ as the optimized parameters for the driving field.
	
	\section{Gaussian pulses}\label{CIN}	
	
As a specific example, we consider the case that the bosonic mode is driven by a series of Gaussian pulses in this section, and the envelope of the driving field is written as
\begin{equation}
	\varepsilon (t) = \frac{\varepsilon _p A }{\sqrt{\pi}}\sum_{m}\exp \left[ - A^{2} \gamma^{2} (t-mT)^{2}\right],
\end{equation}	
where $\varepsilon _p$ denotes the driving strength, $A$ governs the duration of the pulses, 
$T$ represents the period of the pulses and $m$ is an integer.
In order to optimize the dynamic blockade, we search for the minimum value of $g^{(2)}\left( t\right) $ in the parameters space of the pulses ($\Delta$, $\varepsilon _p$, $T$, and $A$) based on the PSO algorithm with $20$ particles and $50$ iterations. The ranges of these parameters for optimization are set as follows: $\Delta/\gamma$ in the range $(-5,5)$, $\varepsilon _p/\gamma$ in the range $(0.1,0.5)$, $\gamma T$ in the range $(3, 8)$, and $A$ in the range $(0.001,10)$. 
	
	\begin{figure}[tbp]
		\centering
		\includegraphics[bb=60 150 2300 2000, width=8.5 cm, clip]{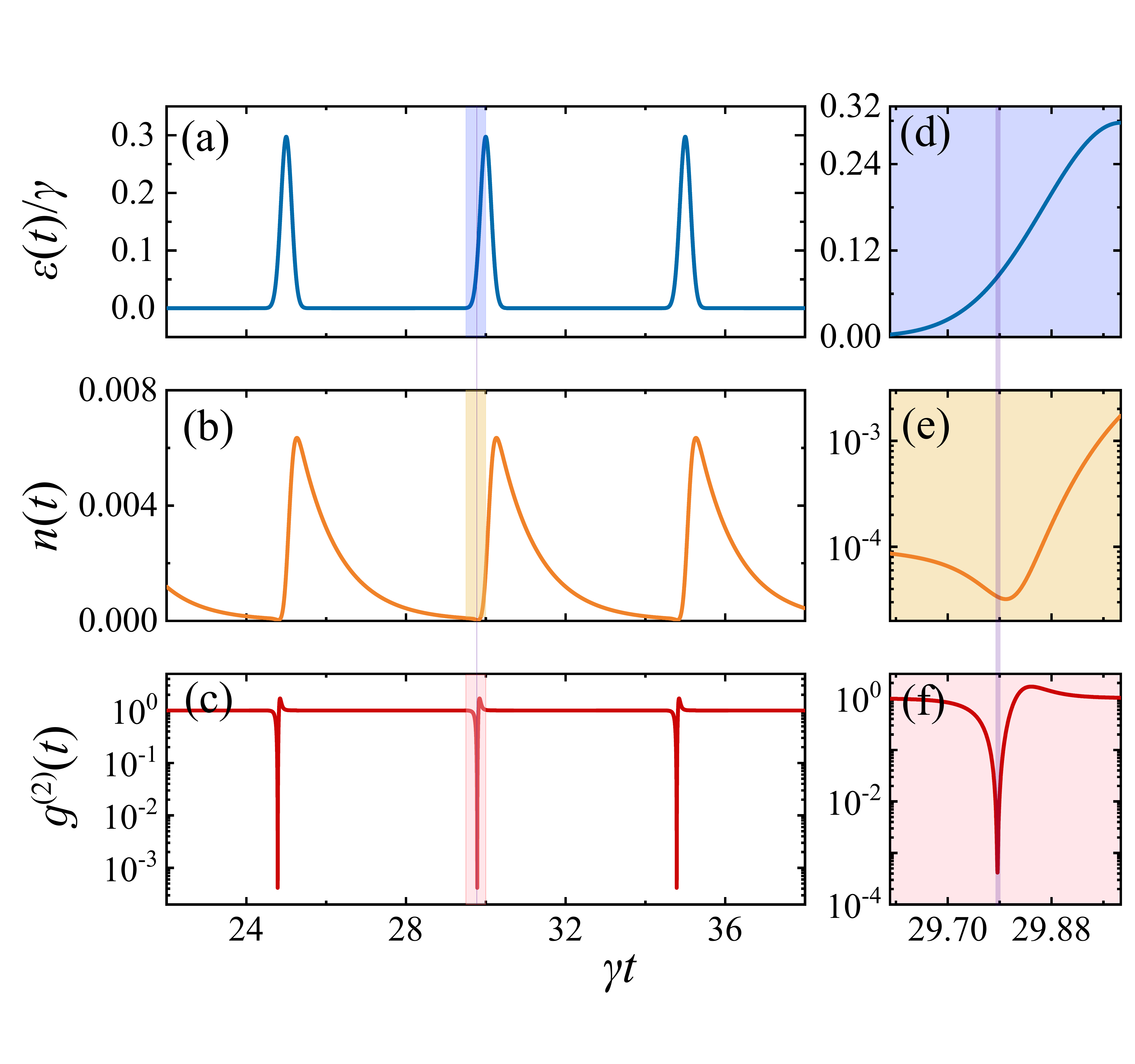}
		\caption{(Color online) (a) The envelope of the driving Gaussian pulses $\varepsilon (t) $, (b) the mean photon number $n(t)$, and (c) the equal-time second-order correlation function  $g^{(2)}\left( t\right)$ versus time $\gamma t$.
                 (d), (e), and (f) are the local enlarged view of (a), (b), and (c), respectively.
			The parameters are $U/\gamma=0.05$, $\gamma T=5$, $\Delta/\gamma=0.5$, $\varepsilon _p/\gamma=0.1$, and $A=5.27$.}   
		\label{fig1}
	\end{figure}

After optimization, we obtain the following optimal parameters for dynamic blockade: $\Delta /\gamma=0.5$, $\varepsilon _p/\gamma=0.1$, $\gamma T=5$, and $A=5.27$. The envelope of the pulses $\varepsilon (t)$ for the optimized parameters is shown in Fig.~\ref{fig1}(a), and the corresponding mean photon number $n(t)$ and second-order correlation function $g^{(2)}(t)$ are shown in Figs.~\ref{fig1}(b) and \ref{fig1}(c), respectively. 
The bosonic mode is periodically excited by the driving pulses, and there is a time delay between the driving pulses and the excitations [Figs.~\ref{fig1}(d)-\ref{fig1}(f)]. 
Such time delay induces a counterintuitive phenomenon that the decay of mean photon number is accelerated and the minimal mean photon number $n_{\rm min}(t)\approx 3\times 10^{-4}$ [Fig.~\ref{fig1}(e)] appears during the increasing of the driving strength [Fig.~\ref{fig1}(d)].
Meanwhile, the second-order correlation function also decays rapidly in this regime and the strong antibunching is obtained with the minimum value $g^{(2)}_{\rm min}\left( t\right)\approx 4\times 10^{-4}$ [Fig.~\ref{fig1}(f)].

	\begin{figure}[tbp]
		\includegraphics[bb=140 250 3400 3300, width=8 cm, clip]{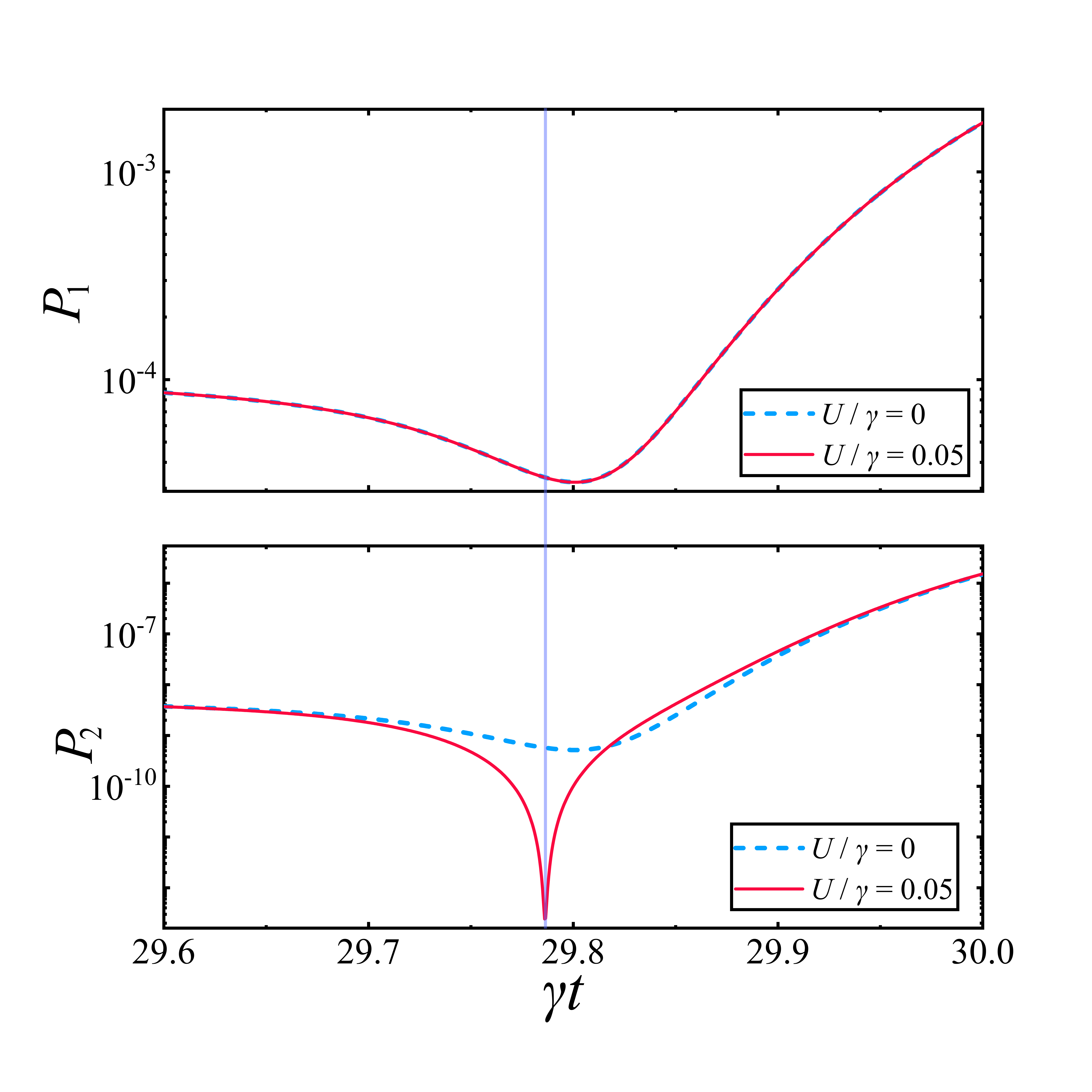}
		\caption{(Color online) The populations of Fock states with $1$ photon and $2$ photons ($P_1$ and $P_2$) versus time $\gamma t$ for $U=0$ (blue dashed curves) and $U/\gamma=0.05$ (red solid curves). The parameters are the same as in Fig.~\ref{fig1}.}   \label{fig2}
	\end{figure}

The populations of Fock states with $1$ photon and $2$ photons ($P_1$ and $P_2$) are shown in Fig.~\ref{fig2}.
If there is no nonlinear interactions ($U=0$) in the bosonic mode, the dynamic of $P_1$ and $P_2$ are simultaneous, and there is no blockade effect. Under the weak nonlinear interaction ($U=0.05\gamma$), the dynamic of $P_1$ and $P_2$ become nonsimultaneous. Most importantly, the minimum value of $P_2$ is about four orders smaller than that for $U=0$ due to the destructive interference between the paths for two-photon excitation, which is the origin of the strong antibunching effect in the weak nonlinear regime.

	\begin{figure}[tbp]
		\includegraphics[bb=30 200 3400 3350, width=7.5 cm, clip]{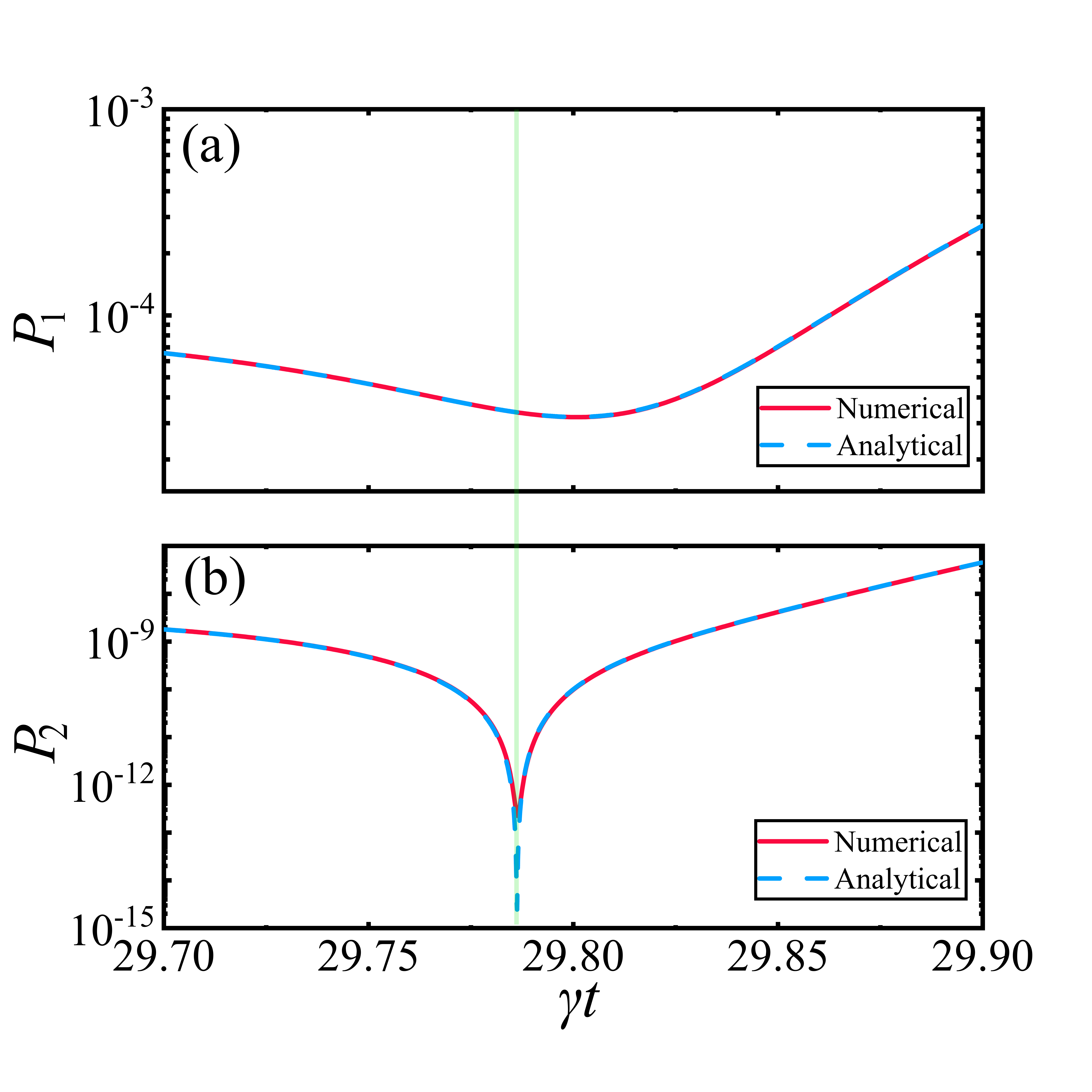}
		\caption{(Color online) The populations of Fock states $|1\rangle$ and $|2\rangle$ ($P_1$ and $P_2$) versus time $\gamma t$ by the numerical [Eq.~(\ref{MasterEq})] (red solid curves) and analytical [Eqs.~(\ref{eq_c1}) and (\ref{eq_c2})] (blue dashed curves) methods. The parameters are the same as in Fig.~\ref{fig1}.}   \label{fig3}
	\end{figure}

To understand the interference between different paths for two-photon excitation, we derive the populations of Fock states based on the Schr\"{o}dinger equation.
Under the weak excitation condition $n(t)\ll 1$, the state of the bosonic mode can be truncated to the two-photon manifold as
\begin{equation}
  | \psi \rangle \approx c_0 |0 \rangle+ c_1 |1 \rangle+ c_2 |2 \rangle.
\end{equation}
Here, $|n \rangle$ is the Fock state with $n$ photons, and $c_n$ is the corresponding coefficient that satisfies the condition $|c_0| \approx 1 \gg |c_1| \gg |c_2|$.
Based on the Schr\"{o}dinger equation, $i \partial | \psi \rangle/\partial t=H_{\rm eff}| \psi \rangle$, with the effective Hamiltonian $H_{\rm eff}\equiv H-i\gamma a^{\dagger}a/2$, the dynamic equations for the coefficients $c_1$ and $c_2$ are given by
\begin{eqnarray}
  \frac{dc_1}{dt} &=& (-i\Delta-\frac{\gamma}{2})c_1 -i\varepsilon(t), \\
  \frac{dc_2}{dt} &=& [-i2(\Delta+U)-\gamma]c_2 -i\sqrt{2}\varepsilon(t)c_1.
\end{eqnarray}
The envelop of the pulsed driving field can be written as a Fourier series as 
\begin{equation}
  \varepsilon(t)= \sum ^{+\infty}_{k=-\infty} \varepsilon_k \exp\left(ik \omega_p t\right),
\end{equation}
where the complex coefficients are given by
\begin{equation}
  \varepsilon_k=\frac{1}{T}\int^{T}_0 \varepsilon(t) \exp\left(-ik\omega_p t\right) dt.
\end{equation}
$\omega_p\equiv 2\pi/T$ is the generation frequency of the Gaussian shaped pulses, and $k$ is an integer.
The dynamic equations can be solved via Fourier transformation as
\begin{eqnarray}
  c_{1} &=& \sum_{k=-\infty }^{+\infty }\chi _{k}^{(1)}e^{ik\omega _{p}t},\label{eq_c1} \\
  c_{2} &=& \sum_{k^{\prime }=-\infty }^{+\infty }\sum_{k=-\infty }^{+\infty
}\chi _{k^{\prime }k}^{(2)}\chi _{k}^{(1)}e^{i(k+k^{\prime
})\omega _{p}t}, \label{eq_c2}
\end{eqnarray}
where
\begin{eqnarray}
  \chi _{k}^{(1)} &=& \frac{-i\varepsilon _{k}}{i(\Delta +k\omega _{p})+\frac{%
\gamma }{2}},\\
  \chi _{k^{\prime }k}^{(2)} &=& \frac{-i\sqrt{2}\varepsilon _{k^{\prime }}}{[i(k+k^{\prime
})\omega _{p}+2i\left( \Delta +U\right) +\gamma ]}.
\end{eqnarray}
Based on the analytical solutions, the populations of Fock states $|1\rangle$ and $|2\rangle$ are given by $P_1=|c_1|^2$ and $P_2=|c_2|^2$.

From Eq.~(\ref{eq_c2}), we can see that the two-photon excitation can be achieved through multiple paths, i.e., absorbing two
photons with the same frequency ($k=k'$) or with different frequencies ($k\neq k'$).
All the paths for two-photon excitation can result in destructive interference and lead to the vanishing of the population
of Fock state $|2\rangle$ with $c_2\approx 0$.
The populations of Fock states $|1\rangle$ and $|2\rangle$ ($P_1$ and $P_2$) obtained by numerical [Eq.~(\ref{MasterEq})] (red solid curves) and analytical [Eqs.~(\ref{eq_c1}) and (\ref{eq_c2})] (blue dashed curves) methods are shown in Fig.~\ref{fig3}.
The results obtained by these two methods match quantitatively.


	\begin{figure}[tbp]
		\includegraphics[bb=87 150 3800 3353, width=8 cm, clip]{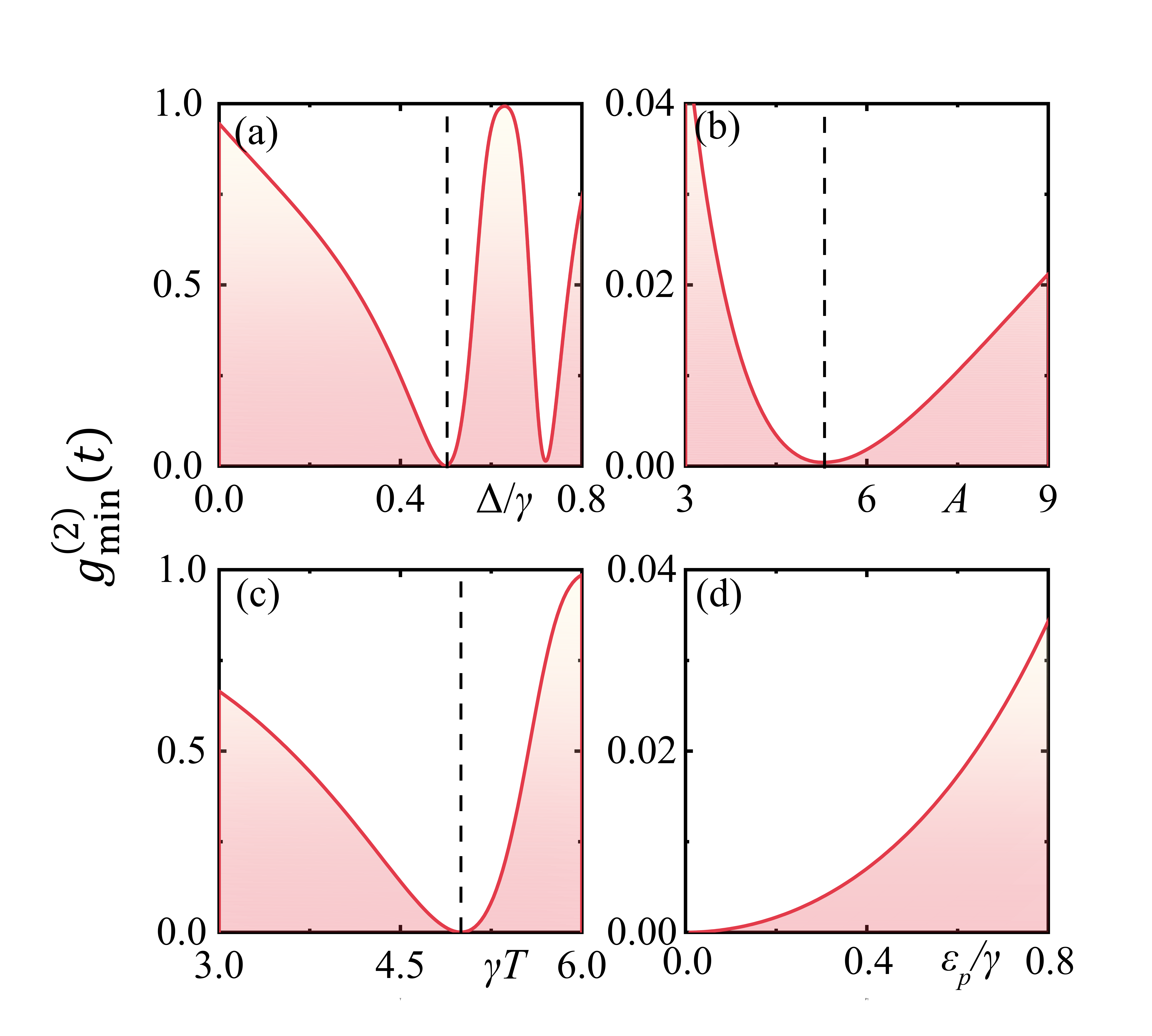}
		\caption{(Color online) The minimal values of the second-order correlation function $g^{(2)}_{\rm min}(t)$ versus parameters: (a) $\Delta /\gamma$, (b) $A$, (c) $\gamma T$, and (d) $\varepsilon _p/\gamma$. The parameters are the same as in Fig.~\ref{fig1}.}   \label{fig4}
	\end{figure}

In order to demonstrate the optimization based on the PSO algorithm, we show the minimal values of the second-order correlation function $g^{(2)}_{\rm min}(t)$ versus the parameters of the driving pulses in Fig.~\ref{fig4}.
From these figures, we can see that the parameters ($\Delta /\gamma=0.5$, $\gamma T=5$, and $A=5.27$) are indeed the optimal parameters for dynamic blockade, which indicates that the PSO algorithm can be used to obtain the optimal conditions for the dynamic blockade.
In addition, there are some notes that should be mentioned: (i) As shown in Fig.~\ref{fig4}(d), a smaller $\varepsilon _p$ corresponds to a smaller $g^{(2)}_{\rm min}\left( t\right)$, as well as a smaller mean photon number $n(t)$.
(ii) In the optimization, we set the range for $\gamma T$ as $(3,8)$, because it is related to the mean photon number $n(t)$.  A longer interval between two pulses results in a smaller mean photon number for strong dynamic blockade, so we impose an upper limit on $\gamma T$ to prevent a too small mean photon number.
(iii) There are many optimal parameter regimes for strong dynamic blockade, so that we can choose different parameters according to the experimental conditions and application situations.
	
	\section{Rectangular pulses}\label{WBON}

	\begin{figure}[tbp]
		\includegraphics[bb=150 150 2400 2000, width=8.5 cm, clip]{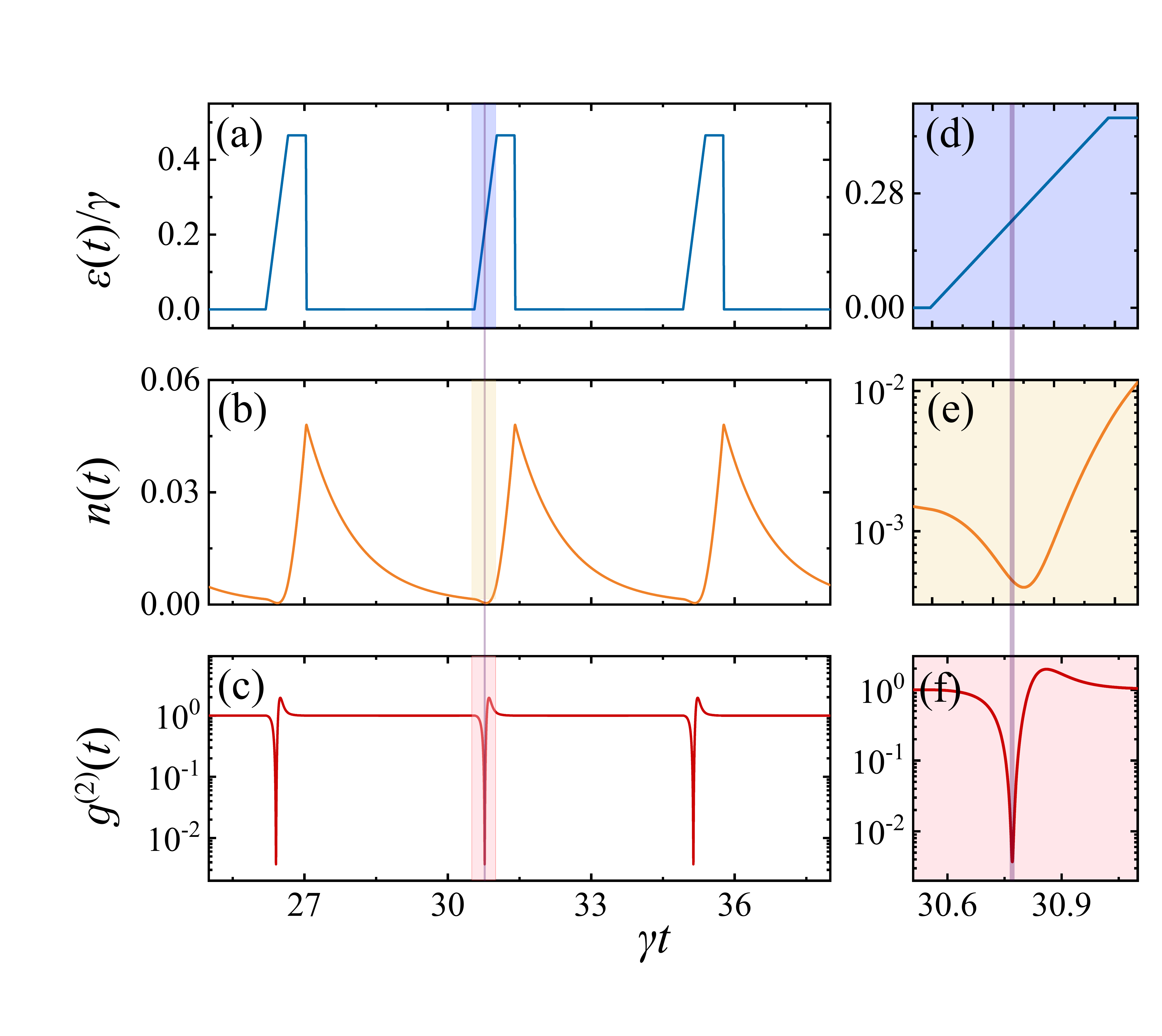}
		\caption{(Color online) (a) The envelope of the driving rectangular pulses $\varepsilon (t) $, (b) the mean photon number $n(t)$, and (c) the equal-time second-order correlation function  $g^{(2)}\left( t\right)$ versus time $\gamma t$.
                 (d), (e), and (f) are the local enlarged view of (a), (b), and (c), respectively.
			The parameters are:  $U/\gamma=0.05$, $\Delta/\gamma=0.617$, $\varepsilon_m/\gamma=0.465$, $\gamma t_r=0.468 $, $\gamma t_w=0.372$, $\gamma t_f=0.016$,  and $\gamma T=4.365$.}   
		\label{fig5}
	\end{figure}

The dynamic blockade optimization based on the PSO algorithm is universal, which also can be applied when the bosonic mode is driven by a pulsed field with different envelope. 
Rectangular pulses are also commonly used in optical driving, and there are even more controllable parameters for the envelope of the rectangular pulses, in comparing with the Gaussian pulses.
In this section, we discuss the dynamic blockade  optimization achieved by driving the bosonic mode with a series of rectangular pulses. 
The envelope of the rectangular pulses is written as a function of time $t$ as
	\begin{equation}
		\varepsilon (t) = 
		\begin{cases} 
			\varepsilon _m \frac{ t'}{t_r},  & 0 \leq t' < t_r, \\
			\varepsilon _m,  &  t_r \leq t' < t_{2}, \\
			\varepsilon _m \frac{(t_{3} - t')}{t_f}, &  t_{2} \leq t' < t_{3},\\
			0, & t_{3} < t' < T,
		\end{cases}
	\end{equation}
where $\varepsilon_m$ represents the maximal amplitude of the rectangular pulses, $t_r$ designates the pulse rise time, $t_f$ corresponds to the pulse fall time, $t_w$ signifies the pulse width, $T$ stands for the pulse period, $t_{2} \equiv t_r + t_w $, $t_{3} \equiv  t_r + t_w + t_f$, $t' \equiv t\% T$, and $\%$ is the modulo operation.

\begin{figure}[tbp]
\includegraphics[bb=50 115 2250 2030, width=8.5 cm, clip]{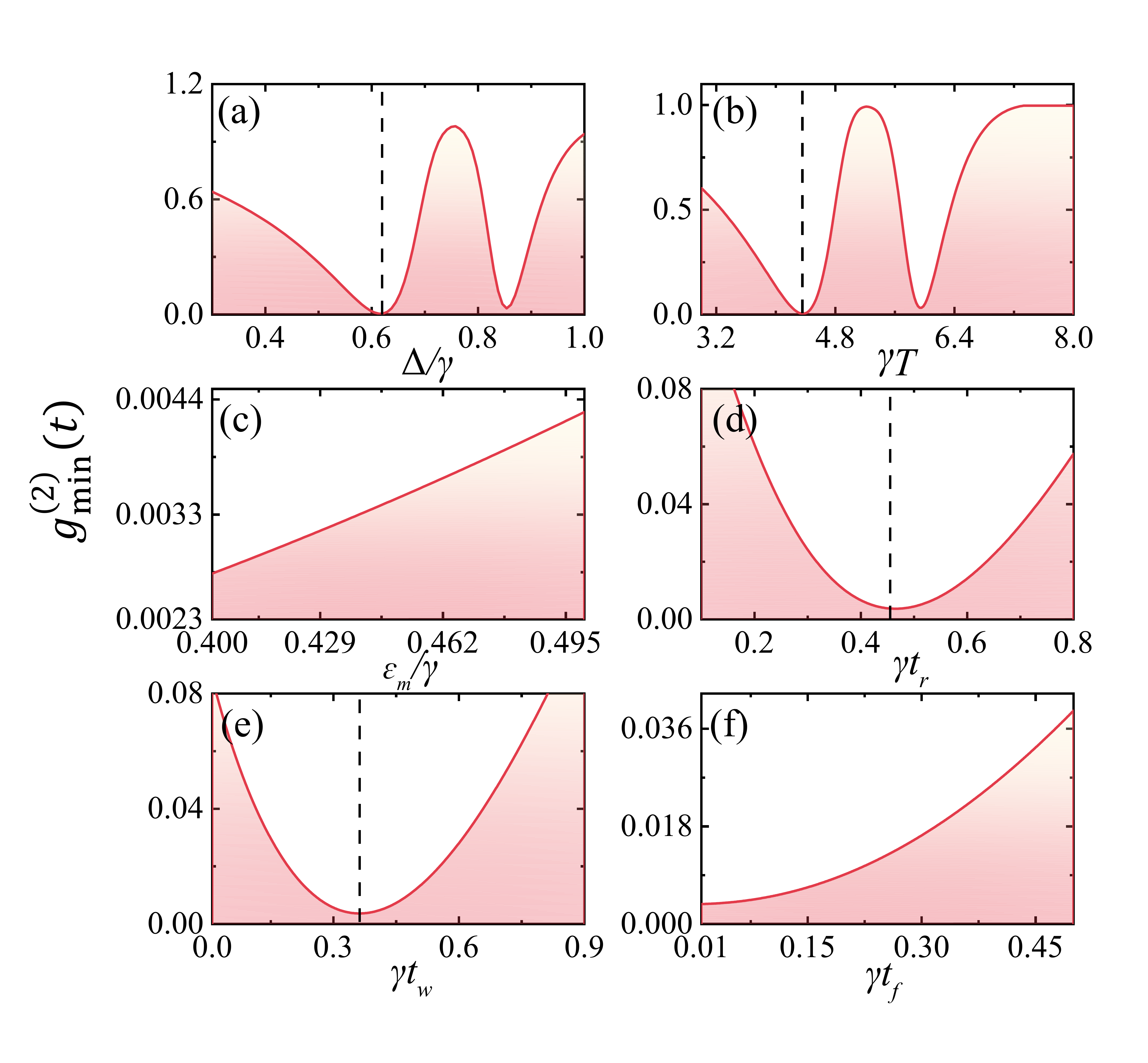}
\caption{(Color online) The minimal values of the second-order correlation function $g^{(2)}_{\rm min}(t)$ versus parameters: (a) $\Delta /\gamma$, (b) $\gamma T$, (c) $\varepsilon_m/\gamma$,  (d) $\gamma t_r$,  (e) $\gamma t_w$, and (f) $\gamma t_f$. The parameters are the same as in Fig.~\ref{fig5}.}   \label{fig6}
\end{figure}

In the optimization based on the PSO algorithm with 20 particles, we set the range of the parameters as: $\Delta/\gamma$ in the range $(-5,5)$, $\varepsilon_m/\gamma$ in the range $(0.1,0.5)$, $\gamma t_r$ in the range $(0.01,0.5)$, $\gamma t_f$ in the range $(0.01,0.5)$, $\gamma t_w$ in the range $(0.01,0.5)$, and $\gamma T$ in the range $(3,8)$. The fitness value has stabilized after 50 iterations, yielding the optimal parameters for this iteration: $\Delta/\gamma=0.617$, $\varepsilon_m/\gamma=0.465$, $\gamma t_r=0.468$, $\gamma t_w=0.372$, $\gamma t_f=0.016$, and $\gamma T=4.365$. 
The envelope of the rectangular pulses $\varepsilon(t)$, the mean photon number $n(t)$, and the second-order correlation function $g^{(2)}\left( t\right)$ plotted as functions of time $t$ with the optimal parameters are shown in Fig.~\ref{fig5}.
The minimal mean photon number ($\approx 4\times 10^{-4}$) [Fig.~\ref{fig5}(e)] is achieved at the rising edges of the rectangular pulses [Fig.~\ref{fig5}(d)], and the dynamic blockade with a minimal value of $g^{(2)}_{\rm min}\left( t\right)\approx 4 \times 10^{-3}$ [Fig.~\ref{fig5}(f)] is obtained just before the mean photon number reaching its minimum value.
As mentioned in the case of Gaussian pulses, the dynamic blockade in the weak nonlinear bosonic mode with pulsed driving is induced by the destructive interference between different paths for two-photon excitation [see Eq.~(\ref{eq_c2})].

To demonstrate the results of optimization, we show the minimal values of the second-order correlation function $g^{(2)}_{\rm min}\left( t\right)$ versus one of the parameters, with the other optimal parameters fixed, in Fig.~\ref{fig6}.
It should be noted that there is no optimal values of both $\varepsilon_m$ and $t_f$, i.e., a smaller $g^{(2)}_{\rm min}\left( t\right)$ is obtained with a smaller $\varepsilon_m$ and $t_f$. Moreover, there are many minimal values in Figs.~\ref{fig6}(a) and \ref{fig6}(b), so, besides the parameters in Fig.~\ref{fig5}, there are some other parameter regimes that also can be used to achieve dynamic blockade.

\section{Conclusions}\label{Con}
	
In conclusion, we have demonstrated dynamic blockade in a single bosonic mode with weak Kerr nonlinearity using only pulsed excitations.
We have proposed a scheme to optimize the envelopes of the driving pulses to generate strong photon blockade by the PSO algorithm.
Based on the analytical expression of the populations of Fock states, we found that there are many paths for two-photon excitation, and the destructive interference between them induces the dynamic blockade in the weak nonlinear regime.
The dynamic blockade optimization based on the PSO algorithm is universal, which has been applied in the cases of driving the single bosonic mode with Gaussian or rectangular pulses. 
Our work opens a way to generate dynamic blockade by the optimization algorithm, which can be generalized to obtain the optimized parameters for observing other quantum effects, such as quantum entanglement and quantum squeezing.

	
	\begin{acknowledgments}
		This work is supported by the
		National Natural Science Foundation of China (NSFC) (Grants No.~12064010 and  No.~12247105), the science and technology innovation Program of Hunan Province (Grant No.~2022RC1203), and Hunan provincial major sci-tech program (Grant No.~2023ZJ1010).
	\end{acknowledgments}

	\bibliography{ref}

\begin{thebibliography}{82}%
\makeatletter
\providecommand \@ifxundefined [1]{%
 \@ifx{#1\undefined}
}%
\providecommand \@ifnum [1]{%
 \ifnum #1\expandafter \@firstoftwo
 \else \expandafter \@secondoftwo
 \fi
}%
\providecommand \@ifx [1]{%
 \ifx #1\expandafter \@firstoftwo
 \else \expandafter \@secondoftwo
 \fi
}%
\providecommand \natexlab [1]{#1}%
\providecommand \enquote  [1]{``#1''}%
\providecommand \bibnamefont  [1]{#1}%
\providecommand \bibfnamefont [1]{#1}%
\providecommand \citenamefont [1]{#1}%
\providecommand \href@noop [0]{\@secondoftwo}%
\providecommand \href [0]{\begingroup \@sanitize@url \@href}%
\providecommand \@href[1]{\@@startlink{#1}\@@href}%
\providecommand \@@href[1]{\endgroup#1\@@endlink}%
\providecommand \@sanitize@url [0]{\catcode `\\12\catcode `\$12\catcode
  `\&12\catcode `\#12\catcode `\^12\catcode `\_12\catcode `\%12\relax}%
\providecommand \@@startlink[1]{}%
\providecommand \@@endlink[0]{}%
\providecommand \url  [0]{\begingroup\@sanitize@url \@url }%
\providecommand \@url [1]{\endgroup\@href {#1}{\urlprefix }}%
\providecommand \urlprefix  [0]{URL }%
\providecommand \Eprint [0]{\href }%
\providecommand \doibase [0]{https://doi.org/}%
\providecommand \selectlanguage [0]{\@gobble}%
\providecommand \bibinfo  [0]{\@secondoftwo}%
\providecommand \bibfield  [0]{\@secondoftwo}%
\providecommand \translation [1]{[#1]}%
\providecommand \BibitemOpen [0]{}%
\providecommand \bibitemStop [0]{}%
\providecommand \bibitemNoStop [0]{.\EOS\space}%
\providecommand \EOS [0]{\spacefactor3000\relax}%
\providecommand \BibitemShut  [1]{\csname bibitem#1\endcsname}%
\let\auto@bib@innerbib\@empty
\bibitem [{\citenamefont {Imamo\ifmmode~\bar{g}\else \={g}\fi{}lu}\ \emph
  {et~al.}(1997)\citenamefont {Imamo\ifmmode~\bar{g}\else \={g}\fi{}lu},
  \citenamefont {Schmidt}, \citenamefont {Woods},\ and\ \citenamefont
  {Deutsch}}]{PhysRevLett.79.1467}%
  \BibitemOpen
  \bibfield  {author} {\bibinfo {author} {\bibfnamefont {A.}~\bibnamefont
  {Imamo\ifmmode~\bar{g}\else \={g}\fi{}lu}}, \bibinfo {author} {\bibfnamefont
  {H.}~\bibnamefont {Schmidt}}, \bibinfo {author} {\bibfnamefont
  {G.}~\bibnamefont {Woods}},\ and\ \bibinfo {author} {\bibfnamefont
  {M.}~\bibnamefont {Deutsch}},\ }\bibfield  {title} {\bibinfo {title}
  {Strongly interacting photons in a nonlinear cavity},\ }\href
  {https://doi.org/10.1103/PhysRevLett.79.1467} {\bibfield  {journal} {\bibinfo
   {journal} {Phys. Rev. Lett.}\ }\textbf {\bibinfo {volume} {79}},\ \bibinfo
  {pages} {1467} (\bibinfo {year} {1997})}\BibitemShut {NoStop}%
\bibitem [{\citenamefont {Buckley}\ \emph {et~al.}(2012)\citenamefont
  {Buckley}, \citenamefont {Rivoire},\ and\ \citenamefont
  {Vučković}}]{Buckley_2012}%
  \BibitemOpen
  \bibfield  {author} {\bibinfo {author} {\bibfnamefont {S.}~\bibnamefont
  {Buckley}}, \bibinfo {author} {\bibfnamefont {K.}~\bibnamefont {Rivoire}},\
  and\ \bibinfo {author} {\bibfnamefont {J.}~\bibnamefont {Vučković}},\
  }\bibfield  {title} {\bibinfo {title} {Engineered quantum dot single-photon
  sources},\ }\href {https://doi.org/10.1088/0034-4885/75/12/126503} {\bibfield
   {journal} {\bibinfo  {journal} {Rep. Prog. Phys.}\ }\textbf {\bibinfo
  {volume} {75}},\ \bibinfo {pages} {126503} (\bibinfo {year}
  {2012})}\BibitemShut {NoStop}%
\bibitem [{\citenamefont {Lodahl}\ \emph {et~al.}(2015)\citenamefont {Lodahl},
  \citenamefont {Mahmoodian},\ and\ \citenamefont
  {Stobbe}}]{RevModPhys.87.347}%
  \BibitemOpen
  \bibfield  {author} {\bibinfo {author} {\bibfnamefont {P.}~\bibnamefont
  {Lodahl}}, \bibinfo {author} {\bibfnamefont {S.}~\bibnamefont {Mahmoodian}},\
  and\ \bibinfo {author} {\bibfnamefont {S.}~\bibnamefont {Stobbe}},\
  }\bibfield  {title} {\bibinfo {title} {Interfacing single photons and single
  quantum dots with photonic nanostructures},\ }\href
  {https://doi.org/10.1103/RevModPhys.87.347} {\bibfield  {journal} {\bibinfo
  {journal} {Rev. Mod. Phys.}\ }\textbf {\bibinfo {volume} {87}},\ \bibinfo
  {pages} {347} (\bibinfo {year} {2015})}\BibitemShut {NoStop}%
\bibitem [{\citenamefont {Knill}\ \emph {et~al.}(2001)\citenamefont {Knill},
  \citenamefont {Laflamme},\ and\ \citenamefont {Milburn}}]{RN565}%
  \BibitemOpen
  \bibfield  {author} {\bibinfo {author} {\bibfnamefont {E.}~\bibnamefont
  {Knill}}, \bibinfo {author} {\bibfnamefont {R.}~\bibnamefont {Laflamme}},\
  and\ \bibinfo {author} {\bibfnamefont {G.~J.}\ \bibnamefont {Milburn}},\
  }\bibfield  {title} {\bibinfo {title} {A scheme for efficient quantum
  computation with linear optics},\ }\href {https://doi.org/10.1038/35051009}
  {\bibfield  {journal} {\bibinfo  {journal} {\nat}\ }\textbf {\bibinfo
  {volume} {409}},\ \bibinfo {pages} {46} (\bibinfo {year} {2001})}\BibitemShut
  {NoStop}%
\bibitem [{\citenamefont {Kok}\ \emph {et~al.}(2007)\citenamefont {Kok},
  \citenamefont {Munro}, \citenamefont {Nemoto}, \citenamefont {Ralph},
  \citenamefont {Dowling},\ and\ \citenamefont {Milburn}}]{RevModPhys.79.135}%
  \BibitemOpen
  \bibfield  {author} {\bibinfo {author} {\bibfnamefont {P.}~\bibnamefont
  {Kok}}, \bibinfo {author} {\bibfnamefont {W.~J.}\ \bibnamefont {Munro}},
  \bibinfo {author} {\bibfnamefont {K.}~\bibnamefont {Nemoto}}, \bibinfo
  {author} {\bibfnamefont {T.~C.}\ \bibnamefont {Ralph}}, \bibinfo {author}
  {\bibfnamefont {J.~P.}\ \bibnamefont {Dowling}},\ and\ \bibinfo {author}
  {\bibfnamefont {G.~J.}\ \bibnamefont {Milburn}},\ }\bibfield  {title}
  {\bibinfo {title} {Linear optical quantum computing with photonic qubits},\
  }\href {https://doi.org/10.1103/RevModPhys.79.135} {\bibfield  {journal}
  {\bibinfo  {journal} {Rev. Mod. Phys.}\ }\textbf {\bibinfo {volume} {79}},\
  \bibinfo {pages} {135} (\bibinfo {year} {2007})}\BibitemShut {NoStop}%
\bibitem [{\citenamefont {Kimble}(2008)}]{RN566}%
  \BibitemOpen
  \bibfield  {author} {\bibinfo {author} {\bibfnamefont {H.~J.}\ \bibnamefont
  {Kimble}},\ }\bibfield  {title} {\bibinfo {title} {The quantum internet},\
  }\href {https://doi.org/10.1038/nature07127} {\bibfield  {journal} {\bibinfo
  {journal} {\nat}\ }\textbf {\bibinfo {volume} {453}},\ \bibinfo {pages}
  {1023} (\bibinfo {year} {2008})}\BibitemShut {NoStop}%
\bibitem [{\citenamefont {Reiserer}\ and\ \citenamefont
  {Rempe}(2015)}]{RevModPhys.87.1379}%
  \BibitemOpen
  \bibfield  {author} {\bibinfo {author} {\bibfnamefont {A.}~\bibnamefont
  {Reiserer}}\ and\ \bibinfo {author} {\bibfnamefont {G.}~\bibnamefont
  {Rempe}},\ }\bibfield  {title} {\bibinfo {title} {Cavity-based quantum
  networks with single atoms and optical photons},\ }\href
  {https://doi.org/10.1103/RevModPhys.87.1379} {\bibfield  {journal} {\bibinfo
  {journal} {Rev. Mod. Phys.}\ }\textbf {\bibinfo {volume} {87}},\ \bibinfo
  {pages} {1379} (\bibinfo {year} {2015})}\BibitemShut {NoStop}%
\bibitem [{\citenamefont {Scarani}\ \emph {et~al.}(2009)\citenamefont
  {Scarani}, \citenamefont {Bechmann-Pasquinucci}, \citenamefont {Cerf},
  \citenamefont {Du\ifmmode~\check{s}\else \v{s}\fi{}ek}, \citenamefont
  {L\"utkenhaus},\ and\ \citenamefont {Peev}}]{RevModPhys.81.1301}%
  \BibitemOpen
  \bibfield  {author} {\bibinfo {author} {\bibfnamefont {V.}~\bibnamefont
  {Scarani}}, \bibinfo {author} {\bibfnamefont {H.}~\bibnamefont
  {Bechmann-Pasquinucci}}, \bibinfo {author} {\bibfnamefont {N.~J.}\
  \bibnamefont {Cerf}}, \bibinfo {author} {\bibfnamefont {M.}~\bibnamefont
  {Du\ifmmode~\check{s}\else \v{s}\fi{}ek}}, \bibinfo {author} {\bibfnamefont
  {N.}~\bibnamefont {L\"utkenhaus}},\ and\ \bibinfo {author} {\bibfnamefont
  {M.}~\bibnamefont {Peev}},\ }\bibfield  {title} {\bibinfo {title} {The
  security of practical quantum key distribution},\ }\href
  {https://doi.org/10.1103/RevModPhys.81.1301} {\bibfield  {journal} {\bibinfo
  {journal} {Rev. Mod. Phys.}\ }\textbf {\bibinfo {volume} {81}},\ \bibinfo
  {pages} {1301} (\bibinfo {year} {2009})}\BibitemShut {NoStop}%
\bibitem [{\citenamefont {Xu}\ \emph {et~al.}(2020)\citenamefont {Xu},
  \citenamefont {Ma}, \citenamefont {Zhang}, \citenamefont {Lo},\ and\
  \citenamefont {Pan}}]{RevModPhys.92.025002}%
  \BibitemOpen
  \bibfield  {author} {\bibinfo {author} {\bibfnamefont {F.}~\bibnamefont
  {Xu}}, \bibinfo {author} {\bibfnamefont {X.}~\bibnamefont {Ma}}, \bibinfo
  {author} {\bibfnamefont {Q.}~\bibnamefont {Zhang}}, \bibinfo {author}
  {\bibfnamefont {H.-K.}\ \bibnamefont {Lo}},\ and\ \bibinfo {author}
  {\bibfnamefont {J.-W.}\ \bibnamefont {Pan}},\ }\bibfield  {title} {\bibinfo
  {title} {Secure quantum key distribution with realistic devices},\ }\href
  {https://doi.org/10.1103/RevModPhys.92.025002} {\bibfield  {journal}
  {\bibinfo  {journal} {Rev. Mod. Phys.}\ }\textbf {\bibinfo {volume} {92}},\
  \bibinfo {pages} {025002} (\bibinfo {year} {2020})}\BibitemShut {NoStop}%
\bibitem [{\citenamefont {Degen}\ \emph {et~al.}(2017)\citenamefont {Degen},
  \citenamefont {Reinhard},\ and\ \citenamefont
  {Cappellaro}}]{RevModPhys.89.035002}%
  \BibitemOpen
  \bibfield  {author} {\bibinfo {author} {\bibfnamefont {C.~L.}\ \bibnamefont
  {Degen}}, \bibinfo {author} {\bibfnamefont {F.}~\bibnamefont {Reinhard}},\
  and\ \bibinfo {author} {\bibfnamefont {P.}~\bibnamefont {Cappellaro}},\
  }\bibfield  {title} {\bibinfo {title} {Quantum sensing},\ }\href
  {https://doi.org/10.1103/RevModPhys.89.035002} {\bibfield  {journal}
  {\bibinfo  {journal} {Rev. Mod. Phys.}\ }\textbf {\bibinfo {volume} {89}},\
  \bibinfo {pages} {035002} (\bibinfo {year} {2017})}\BibitemShut {NoStop}%
\bibitem [{\citenamefont {{Pirandola}}\ \emph {et~al.}(2018)\citenamefont
  {{Pirandola}}, \citenamefont {{Bardhan}}, \citenamefont {{Gehring}},
  \citenamefont {{Weedbrook}},\ and\ \citenamefont
  {{Lloyd}}}]{Pirandola2018NaPho}%
  \BibitemOpen
  \bibfield  {author} {\bibinfo {author} {\bibfnamefont {S.}~\bibnamefont
  {{Pirandola}}}, \bibinfo {author} {\bibfnamefont {B.~R.}\ \bibnamefont
  {{Bardhan}}}, \bibinfo {author} {\bibfnamefont {T.}~\bibnamefont
  {{Gehring}}}, \bibinfo {author} {\bibfnamefont {C.}~\bibnamefont
  {{Weedbrook}}},\ and\ \bibinfo {author} {\bibfnamefont {S.}~\bibnamefont
  {{Lloyd}}},\ }\bibfield  {title} {\bibinfo {title} {{Advances in photonic
  quantum sensing}},\ }\href {https://doi.org/10.1038/s41566-018-0301-6}
  {\bibfield  {journal} {\bibinfo  {journal} {Nature Photon.}\ }\textbf
  {\bibinfo {volume} {12}},\ \bibinfo {pages} {724} (\bibinfo {year}
  {2018})}\BibitemShut {NoStop}%
\bibitem [{\citenamefont {Ridolfo}\ \emph {et~al.}(2012)\citenamefont
  {Ridolfo}, \citenamefont {Leib}, \citenamefont {Savasta},\ and\ \citenamefont
  {Hartmann}}]{Ridolfo2012PRL}%
  \BibitemOpen
  \bibfield  {author} {\bibinfo {author} {\bibfnamefont {A.}~\bibnamefont
  {Ridolfo}}, \bibinfo {author} {\bibfnamefont {M.}~\bibnamefont {Leib}},
  \bibinfo {author} {\bibfnamefont {S.}~\bibnamefont {Savasta}},\ and\ \bibinfo
  {author} {\bibfnamefont {M.~J.}\ \bibnamefont {Hartmann}},\ }\bibfield
  {title} {\bibinfo {title} {Photon blockade in the ultrastrong coupling
  regime},\ }\href {https://doi.org/10.1103/PhysRevLett.109.193602} {\bibfield
  {journal} {\bibinfo  {journal} {Phys. Rev. Lett.}\ }\textbf {\bibinfo
  {volume} {109}},\ \bibinfo {pages} {193602} (\bibinfo {year}
  {2012})}\BibitemShut {NoStop}%
\bibitem [{\citenamefont {Liu}\ \emph {et~al.}(2014)\citenamefont {Liu},
  \citenamefont {Xu}, \citenamefont {Miranowicz},\ and\ \citenamefont
  {Nori}}]{LiuYX2014PRA}%
  \BibitemOpen
  \bibfield  {author} {\bibinfo {author} {\bibfnamefont {Y.-x.}\ \bibnamefont
  {Liu}}, \bibinfo {author} {\bibfnamefont {X.-W.}\ \bibnamefont {Xu}},
  \bibinfo {author} {\bibfnamefont {A.}~\bibnamefont {Miranowicz}},\ and\
  \bibinfo {author} {\bibfnamefont {F.}~\bibnamefont {Nori}},\ }\bibfield
  {title} {\bibinfo {title} {From blockade to transparency: Controllable photon
  transmission through a circuit-qed system},\ }\href
  {https://doi.org/10.1103/PhysRevA.89.043818} {\bibfield  {journal} {\bibinfo
  {journal} {Phys. Rev. A}\ }\textbf {\bibinfo {volume} {89}},\ \bibinfo
  {pages} {043818} (\bibinfo {year} {2014})}\BibitemShut {NoStop}%
\bibitem [{\citenamefont {Miranowicz}\ \emph {et~al.}(2014)\citenamefont
  {Miranowicz}, \citenamefont {Bajer}, \citenamefont {Paprzycka}, \citenamefont
  {Liu}, \citenamefont {Zagoskin},\ and\ \citenamefont {Nori}}]{Adam2014PRA}%
  \BibitemOpen
  \bibfield  {author} {\bibinfo {author} {\bibfnamefont {A.}~\bibnamefont
  {Miranowicz}}, \bibinfo {author} {\bibfnamefont {J.}~\bibnamefont {Bajer}},
  \bibinfo {author} {\bibfnamefont {M.}~\bibnamefont {Paprzycka}}, \bibinfo
  {author} {\bibfnamefont {Y.-x.}\ \bibnamefont {Liu}}, \bibinfo {author}
  {\bibfnamefont {A.~M.}\ \bibnamefont {Zagoskin}},\ and\ \bibinfo {author}
  {\bibfnamefont {F.}~\bibnamefont {Nori}},\ }\bibfield  {title} {\bibinfo
  {title} {State-dependent photon blockade via quantum-reservoir engineering},\
  }\href {https://doi.org/10.1103/PhysRevA.90.033831} {\bibfield  {journal}
  {\bibinfo  {journal} {Phys. Rev. A}\ }\textbf {\bibinfo {volume} {90}},\
  \bibinfo {pages} {033831} (\bibinfo {year} {2014})}\BibitemShut {NoStop}%
\bibitem [{\citenamefont {Rabl}(2011)}]{PhysRevLett.107.063601}%
  \BibitemOpen
  \bibfield  {author} {\bibinfo {author} {\bibfnamefont {P.}~\bibnamefont
  {Rabl}},\ }\bibfield  {title} {\bibinfo {title} {Photon blockade effect in
  optomechanical systems},\ }\href
  {https://doi.org/10.1103/PhysRevLett.107.063601} {\bibfield  {journal}
  {\bibinfo  {journal} {Phys. Rev. Lett.}\ }\textbf {\bibinfo {volume} {107}},\
  \bibinfo {pages} {063601} (\bibinfo {year} {2011})}\BibitemShut {NoStop}%
\bibitem [{\citenamefont {Nunnenkamp}\ \emph {et~al.}(2011)\citenamefont
  {Nunnenkamp}, \citenamefont {B\o{}rkje},\ and\ \citenamefont
  {Girvin}}]{PhysRevLett.107.063602}%
  \BibitemOpen
  \bibfield  {author} {\bibinfo {author} {\bibfnamefont {A.}~\bibnamefont
  {Nunnenkamp}}, \bibinfo {author} {\bibfnamefont {K.}~\bibnamefont
  {B\o{}rkje}},\ and\ \bibinfo {author} {\bibfnamefont {S.~M.}\ \bibnamefont
  {Girvin}},\ }\bibfield  {title} {\bibinfo {title} {Single-photon
  optomechanics},\ }\href {https://doi.org/10.1103/PhysRevLett.107.063602}
  {\bibfield  {journal} {\bibinfo  {journal} {Phys. Rev. Lett.}\ }\textbf
  {\bibinfo {volume} {107}},\ \bibinfo {pages} {063602} (\bibinfo {year}
  {2011})}\BibitemShut {NoStop}%
\bibitem [{\citenamefont {Liao}\ and\ \citenamefont
  {Nori}(2013)}]{LiaoJQ2013PRA}%
  \BibitemOpen
  \bibfield  {author} {\bibinfo {author} {\bibfnamefont {J.-Q.}\ \bibnamefont
  {Liao}}\ and\ \bibinfo {author} {\bibfnamefont {F.}~\bibnamefont {Nori}},\
  }\bibfield  {title} {\bibinfo {title} {Photon blockade in quadratically
  coupled optomechanical systems},\ }\href
  {https://doi.org/10.1103/PhysRevA.88.023853} {\bibfield  {journal} {\bibinfo
  {journal} {Phys. Rev. A}\ }\textbf {\bibinfo {volume} {88}},\ \bibinfo
  {pages} {023853} (\bibinfo {year} {2013})}\BibitemShut {NoStop}%
\bibitem [{\citenamefont {Xie}\ \emph {et~al.}(2016)\citenamefont {Xie},
  \citenamefont {Lin}, \citenamefont {Chen}, \citenamefont {Chen},\ and\
  \citenamefont {Lin}}]{XieH2016PRA}%
  \BibitemOpen
  \bibfield  {author} {\bibinfo {author} {\bibfnamefont {H.}~\bibnamefont
  {Xie}}, \bibinfo {author} {\bibfnamefont {G.-W.}\ \bibnamefont {Lin}},
  \bibinfo {author} {\bibfnamefont {X.}~\bibnamefont {Chen}}, \bibinfo {author}
  {\bibfnamefont {Z.-H.}\ \bibnamefont {Chen}},\ and\ \bibinfo {author}
  {\bibfnamefont {X.-M.}\ \bibnamefont {Lin}},\ }\bibfield  {title} {\bibinfo
  {title} {Single-photon nonlinearities in a strongly driven optomechanical
  system with quadratic coupling},\ }\href
  {https://doi.org/10.1103/PhysRevA.93.063860} {\bibfield  {journal} {\bibinfo
  {journal} {Phys. Rev. A}\ }\textbf {\bibinfo {volume} {93}},\ \bibinfo
  {pages} {063860} (\bibinfo {year} {2016})}\BibitemShut {NoStop}%
\bibitem [{\citenamefont {Majumdar}\ and\ \citenamefont
  {Gerace}(2013)}]{Majumdar2013PRB}%
  \BibitemOpen
  \bibfield  {author} {\bibinfo {author} {\bibfnamefont {A.}~\bibnamefont
  {Majumdar}}\ and\ \bibinfo {author} {\bibfnamefont {D.}~\bibnamefont
  {Gerace}},\ }\bibfield  {title} {\bibinfo {title} {Single-photon blockade in
  doubly resonant nanocavities with second-order nonlinearity},\ }\href
  {https://doi.org/10.1103/PhysRevB.87.235319} {\bibfield  {journal} {\bibinfo
  {journal} {Phys. Rev. B}\ }\textbf {\bibinfo {volume} {87}},\ \bibinfo
  {pages} {235319} (\bibinfo {year} {2013})}\BibitemShut {NoStop}%
\bibitem [{\citenamefont {Huang}\ \emph {et~al.}(2018)\citenamefont {Huang},
  \citenamefont {Miranowicz}, \citenamefont {Liao}, \citenamefont {Nori},\ and\
  \citenamefont {Jing}}]{Huang2018PRL}%
  \BibitemOpen
  \bibfield  {author} {\bibinfo {author} {\bibfnamefont {R.}~\bibnamefont
  {Huang}}, \bibinfo {author} {\bibfnamefont {A.}~\bibnamefont {Miranowicz}},
  \bibinfo {author} {\bibfnamefont {J.-Q.}\ \bibnamefont {Liao}}, \bibinfo
  {author} {\bibfnamefont {F.}~\bibnamefont {Nori}},\ and\ \bibinfo {author}
  {\bibfnamefont {H.}~\bibnamefont {Jing}},\ }\bibfield  {title} {\bibinfo
  {title} {Nonreciprocal photon blockade},\ }\href
  {https://doi.org/10.1103/PhysRevLett.121.153601} {\bibfield  {journal}
  {\bibinfo  {journal} {Phys. Rev. Lett.}\ }\textbf {\bibinfo {volume} {121}},\
  \bibinfo {pages} {153601} (\bibinfo {year} {2018})}\BibitemShut {NoStop}%
\bibitem [{\citenamefont {{Huang}}\ \emph {et~al.}(2022)\citenamefont
  {{Huang}}, \citenamefont {{{\"O}zdemir}}, \citenamefont {{Liao}},
  \citenamefont {{Minganti}}, \citenamefont {{Kuang}}, \citenamefont {{Nori}},\
  and\ \citenamefont {{Jing}}}]{Huang2022LPRv}%
  \BibitemOpen
  \bibfield  {author} {\bibinfo {author} {\bibfnamefont {R.}~\bibnamefont
  {{Huang}}}, \bibinfo {author} {\bibfnamefont {{\c{S}}.~K.}\ \bibnamefont
  {{{\"O}zdemir}}}, \bibinfo {author} {\bibfnamefont {J.-Q.}\ \bibnamefont
  {{Liao}}}, \bibinfo {author} {\bibfnamefont {F.}~\bibnamefont {{Minganti}}},
  \bibinfo {author} {\bibfnamefont {L.-M.}\ \bibnamefont {{Kuang}}}, \bibinfo
  {author} {\bibfnamefont {F.}~\bibnamefont {{Nori}}},\ and\ \bibinfo {author}
  {\bibfnamefont {H.}~\bibnamefont {{Jing}}},\ }\bibfield  {title} {\bibinfo
  {title} {{Exceptional Photon Blockade: Engineering Photon Blockade with
  Chiral Exceptional Points}},\ }\href {https://doi.org/10.1002/lpor.202100430}
  {\bibfield  {journal} {\bibinfo  {journal} {Laser Photon. Rev.}\ }\textbf
  {\bibinfo {volume} {16}},\ \bibinfo {eid} {2100430} (\bibinfo {year}
  {2022})}\BibitemShut {NoStop}%
\bibitem [{\citenamefont {{Chakram}}\ \emph {et~al.}(2022)\citenamefont
  {{Chakram}}, \citenamefont {{He}}, \citenamefont {{Dixit}}, \citenamefont
  {{Oriani}}, \citenamefont {{Naik}}, \citenamefont {{Leung}}, \citenamefont
  {{Kwon}}, \citenamefont {{Ma}}, \citenamefont {{Jiang}},\ and\ \citenamefont
  {{Schuster}}}]{Chakram2022NatPh}%
  \BibitemOpen
  \bibfield  {author} {\bibinfo {author} {\bibfnamefont {S.}~\bibnamefont
  {{Chakram}}}, \bibinfo {author} {\bibfnamefont {K.}~\bibnamefont {{He}}},
  \bibinfo {author} {\bibfnamefont {A.~V.}\ \bibnamefont {{Dixit}}}, \bibinfo
  {author} {\bibfnamefont {A.~E.}\ \bibnamefont {{Oriani}}}, \bibinfo {author}
  {\bibfnamefont {R.~K.}\ \bibnamefont {{Naik}}}, \bibinfo {author}
  {\bibfnamefont {N.}~\bibnamefont {{Leung}}}, \bibinfo {author} {\bibfnamefont
  {H.}~\bibnamefont {{Kwon}}}, \bibinfo {author} {\bibfnamefont {W.-L.}\
  \bibnamefont {{Ma}}}, \bibinfo {author} {\bibfnamefont {L.}~\bibnamefont
  {{Jiang}}},\ and\ \bibinfo {author} {\bibfnamefont {D.~I.}\ \bibnamefont
  {{Schuster}}},\ }\bibfield  {title} {\bibinfo {title} {{Multimode photon
  blockade}},\ }\href {https://doi.org/10.1038/s41567-022-01630-y} {\bibfield
  {journal} {\bibinfo  {journal} {Nat. Phys.}\ }\textbf {\bibinfo {volume}
  {18}},\ \bibinfo {pages} {879} (\bibinfo {year} {2022})}\BibitemShut
  {NoStop}%
\bibitem [{\citenamefont {Zhou}\ \emph {et~al.}(2020)\citenamefont {Zhou},
  \citenamefont {Zhang}, \citenamefont {Wu}, \citenamefont {Ye}, \citenamefont
  {Zhang}, \citenamefont {Zou}, \citenamefont {Shen},\ and\ \citenamefont
  {Yang}}]{Zhou2020PRA}%
  \BibitemOpen
  \bibfield  {author} {\bibinfo {author} {\bibfnamefont {Y.~H.}\ \bibnamefont
  {Zhou}}, \bibinfo {author} {\bibfnamefont {X.~Y.}\ \bibnamefont {Zhang}},
  \bibinfo {author} {\bibfnamefont {Q.~C.}\ \bibnamefont {Wu}}, \bibinfo
  {author} {\bibfnamefont {B.~L.}\ \bibnamefont {Ye}}, \bibinfo {author}
  {\bibfnamefont {Z.-Q.}\ \bibnamefont {Zhang}}, \bibinfo {author}
  {\bibfnamefont {D.~D.}\ \bibnamefont {Zou}}, \bibinfo {author} {\bibfnamefont
  {H.~Z.}\ \bibnamefont {Shen}},\ and\ \bibinfo {author} {\bibfnamefont
  {C.-P.}\ \bibnamefont {Yang}},\ }\bibfield  {title} {\bibinfo {title}
  {Conventional photon blockade with a three-wave mixing},\ }\href
  {https://doi.org/10.1103/PhysRevA.102.033713} {\bibfield  {journal} {\bibinfo
   {journal} {Phys. Rev. A}\ }\textbf {\bibinfo {volume} {102}},\ \bibinfo
  {pages} {033713} (\bibinfo {year} {2020})}\BibitemShut {NoStop}%
\bibitem [{\citenamefont {{Birnbaum}}\ \emph {et~al.}(2005)\citenamefont
  {{Birnbaum}}, \citenamefont {{Boca}}, \citenamefont {{Miller}}, \citenamefont
  {{Boozer}}, \citenamefont {{Northup}},\ and\ \citenamefont
  {{Kimble}}}]{Birnbaum2005Natur}%
  \BibitemOpen
  \bibfield  {author} {\bibinfo {author} {\bibfnamefont {K.~M.}\ \bibnamefont
  {{Birnbaum}}}, \bibinfo {author} {\bibfnamefont {A.}~\bibnamefont {{Boca}}},
  \bibinfo {author} {\bibfnamefont {R.}~\bibnamefont {{Miller}}}, \bibinfo
  {author} {\bibfnamefont {A.~D.}\ \bibnamefont {{Boozer}}}, \bibinfo {author}
  {\bibfnamefont {T.~E.}\ \bibnamefont {{Northup}}},\ and\ \bibinfo {author}
  {\bibfnamefont {H.~J.}\ \bibnamefont {{Kimble}}},\ }\bibfield  {title}
  {\bibinfo {title} {{Photon blockade in an optical cavity with one trapped
  atom}},\ }\href {https://doi.org/10.1038/nature03804} {\bibfield  {journal}
  {\bibinfo  {journal} {\nat}\ }\textbf {\bibinfo {volume} {436}},\ \bibinfo
  {pages} {87} (\bibinfo {year} {2005})}\BibitemShut {NoStop}%
\bibitem [{\citenamefont {{Dayan}}\ \emph {et~al.}(2008)\citenamefont
  {{Dayan}}, \citenamefont {{Parkins}}, \citenamefont {{Aoki}}, \citenamefont
  {{Ostby}}, \citenamefont {{Vahala}},\ and\ \citenamefont
  {{Kimble}}}]{Dayan2008Sci}%
  \BibitemOpen
  \bibfield  {author} {\bibinfo {author} {\bibfnamefont {B.}~\bibnamefont
  {{Dayan}}}, \bibinfo {author} {\bibfnamefont {A.~S.}\ \bibnamefont
  {{Parkins}}}, \bibinfo {author} {\bibfnamefont {T.}~\bibnamefont {{Aoki}}},
  \bibinfo {author} {\bibfnamefont {E.~P.}\ \bibnamefont {{Ostby}}}, \bibinfo
  {author} {\bibfnamefont {K.~J.}\ \bibnamefont {{Vahala}}},\ and\ \bibinfo
  {author} {\bibfnamefont {H.~J.}\ \bibnamefont {{Kimble}}},\ }\bibfield
  {title} {\bibinfo {title} {{A Photon Turnstile Dynamically Regulated by One
  Atom}},\ }\href {https://doi.org/10.1126/science.1152261} {\bibfield
  {journal} {\bibinfo  {journal} {Science}\ }\textbf {\bibinfo {volume}
  {319}},\ \bibinfo {pages} {1062} (\bibinfo {year} {2008})}\BibitemShut
  {NoStop}%
\bibitem [{\citenamefont {Aoki}\ \emph {et~al.}(2009)\citenamefont {Aoki},
  \citenamefont {Parkins}, \citenamefont {Alton}, \citenamefont {Regal},
  \citenamefont {Dayan}, \citenamefont {Ostby}, \citenamefont {Vahala},\ and\
  \citenamefont {Kimble}}]{Aoki2009PRL}%
  \BibitemOpen
  \bibfield  {author} {\bibinfo {author} {\bibfnamefont {T.}~\bibnamefont
  {Aoki}}, \bibinfo {author} {\bibfnamefont {A.~S.}\ \bibnamefont {Parkins}},
  \bibinfo {author} {\bibfnamefont {D.~J.}\ \bibnamefont {Alton}}, \bibinfo
  {author} {\bibfnamefont {C.~A.}\ \bibnamefont {Regal}}, \bibinfo {author}
  {\bibfnamefont {B.}~\bibnamefont {Dayan}}, \bibinfo {author} {\bibfnamefont
  {E.}~\bibnamefont {Ostby}}, \bibinfo {author} {\bibfnamefont {K.~J.}\
  \bibnamefont {Vahala}},\ and\ \bibinfo {author} {\bibfnamefont {H.~J.}\
  \bibnamefont {Kimble}},\ }\bibfield  {title} {\bibinfo {title} {Efficient
  routing of single photons by one atom and a microtoroidal cavity},\ }\href
  {https://doi.org/10.1103/PhysRevLett.102.083601} {\bibfield  {journal}
  {\bibinfo  {journal} {Phys. Rev. Lett.}\ }\textbf {\bibinfo {volume} {102}},\
  \bibinfo {pages} {083601} (\bibinfo {year} {2009})}\BibitemShut {NoStop}%
\bibitem [{\citenamefont {{Faraon}}\ \emph {et~al.}(2008)\citenamefont
  {{Faraon}}, \citenamefont {{Fushman}}, \citenamefont {{Englund}},
  \citenamefont {{Stoltz}}, \citenamefont {{Petroff}},\ and\ \citenamefont
  {{Vu{\v{c}}kovi{\'c}}}}]{Faraon2008NatPh}%
  \BibitemOpen
  \bibfield  {author} {\bibinfo {author} {\bibfnamefont {A.}~\bibnamefont
  {{Faraon}}}, \bibinfo {author} {\bibfnamefont {I.}~\bibnamefont {{Fushman}}},
  \bibinfo {author} {\bibfnamefont {D.}~\bibnamefont {{Englund}}}, \bibinfo
  {author} {\bibfnamefont {N.}~\bibnamefont {{Stoltz}}}, \bibinfo {author}
  {\bibfnamefont {P.}~\bibnamefont {{Petroff}}},\ and\ \bibinfo {author}
  {\bibfnamefont {J.}~\bibnamefont {{Vu{\v{c}}kovi{\'c}}}},\ }\bibfield
  {title} {\bibinfo {title} {{Coherent generation of nonclassical light on a
  chip via photon-induced tunneling and blockade}},\ }\href
  {https://doi.org/10.1038/nphys1078} {\bibfield  {journal} {\bibinfo
  {journal} {Nat. Phys.}\ }\textbf {\bibinfo {volume} {4}},\ \bibinfo {pages}
  {859} (\bibinfo {year} {2008})}\BibitemShut {NoStop}%
\bibitem [{\citenamefont {{Reinhard}}\ \emph {et~al.}(2012)\citenamefont
  {{Reinhard}}, \citenamefont {{Volz}}, \citenamefont {{Winger}}, \citenamefont
  {{Badolato}}, \citenamefont {{Hennessy}}, \citenamefont {{Hu}},\ and\
  \citenamefont {{Imamo{\u{g}}lu}}}]{Reinhard2012NaPho}%
  \BibitemOpen
  \bibfield  {author} {\bibinfo {author} {\bibfnamefont {A.}~\bibnamefont
  {{Reinhard}}}, \bibinfo {author} {\bibfnamefont {T.}~\bibnamefont {{Volz}}},
  \bibinfo {author} {\bibfnamefont {M.}~\bibnamefont {{Winger}}}, \bibinfo
  {author} {\bibfnamefont {A.}~\bibnamefont {{Badolato}}}, \bibinfo {author}
  {\bibfnamefont {K.~J.}\ \bibnamefont {{Hennessy}}}, \bibinfo {author}
  {\bibfnamefont {E.~L.}\ \bibnamefont {{Hu}}},\ and\ \bibinfo {author}
  {\bibfnamefont {A.}~\bibnamefont {{Imamo{\u{g}}lu}}},\ }\bibfield  {title}
  {\bibinfo {title} {{Strongly correlated photons on a chip}},\ }\href
  {https://doi.org/10.1038/nphoton.2011.321} {\bibfield  {journal} {\bibinfo
  {journal} {Nature Photon.}\ }\textbf {\bibinfo {volume} {6}},\ \bibinfo
  {pages} {93} (\bibinfo {year} {2012})}\BibitemShut {NoStop}%
\bibitem [{\citenamefont {Lang}\ \emph {et~al.}(2011)\citenamefont {Lang},
  \citenamefont {Bozyigit}, \citenamefont {Eichler}, \citenamefont {Steffen},
  \citenamefont {Fink}, \citenamefont {Abdumalikov}, \citenamefont {Baur},
  \citenamefont {Filipp}, \citenamefont {da~Silva}, \citenamefont {Blais},\
  and\ \citenamefont {Wallraff}}]{LangC2011PRL}%
  \BibitemOpen
  \bibfield  {author} {\bibinfo {author} {\bibfnamefont {C.}~\bibnamefont
  {Lang}}, \bibinfo {author} {\bibfnamefont {D.}~\bibnamefont {Bozyigit}},
  \bibinfo {author} {\bibfnamefont {C.}~\bibnamefont {Eichler}}, \bibinfo
  {author} {\bibfnamefont {L.}~\bibnamefont {Steffen}}, \bibinfo {author}
  {\bibfnamefont {J.~M.}\ \bibnamefont {Fink}}, \bibinfo {author}
  {\bibfnamefont {A.~A.}\ \bibnamefont {Abdumalikov}}, \bibinfo {author}
  {\bibfnamefont {M.}~\bibnamefont {Baur}}, \bibinfo {author} {\bibfnamefont
  {S.}~\bibnamefont {Filipp}}, \bibinfo {author} {\bibfnamefont {M.~P.}\
  \bibnamefont {da~Silva}}, \bibinfo {author} {\bibfnamefont {A.}~\bibnamefont
  {Blais}},\ and\ \bibinfo {author} {\bibfnamefont {A.}~\bibnamefont
  {Wallraff}},\ }\bibfield  {title} {\bibinfo {title} {Observation of resonant
  photon blockade at microwave frequencies using correlation function
  measurements},\ }\href {https://doi.org/10.1103/PhysRevLett.106.243601}
  {\bibfield  {journal} {\bibinfo  {journal} {Phys. Rev. Lett.}\ }\textbf
  {\bibinfo {volume} {106}},\ \bibinfo {pages} {243601} (\bibinfo {year}
  {2011})}\BibitemShut {NoStop}%
\bibitem [{\citenamefont {Hoffman}\ \emph {et~al.}(2011)\citenamefont
  {Hoffman}, \citenamefont {Srinivasan}, \citenamefont {Schmidt}, \citenamefont
  {Spietz}, \citenamefont {Aumentado}, \citenamefont {T\"ureci},\ and\
  \citenamefont {Houck}}]{Hoffman2011PRL}%
  \BibitemOpen
  \bibfield  {author} {\bibinfo {author} {\bibfnamefont {A.~J.}\ \bibnamefont
  {Hoffman}}, \bibinfo {author} {\bibfnamefont {S.~J.}\ \bibnamefont
  {Srinivasan}}, \bibinfo {author} {\bibfnamefont {S.}~\bibnamefont {Schmidt}},
  \bibinfo {author} {\bibfnamefont {L.}~\bibnamefont {Spietz}}, \bibinfo
  {author} {\bibfnamefont {J.}~\bibnamefont {Aumentado}}, \bibinfo {author}
  {\bibfnamefont {H.~E.}\ \bibnamefont {T\"ureci}},\ and\ \bibinfo {author}
  {\bibfnamefont {A.~A.}\ \bibnamefont {Houck}},\ }\bibfield  {title} {\bibinfo
  {title} {Dispersive photon blockade in a superconducting circuit},\ }\href
  {https://doi.org/10.1103/PhysRevLett.107.053602} {\bibfield  {journal}
  {\bibinfo  {journal} {Phys. Rev. Lett.}\ }\textbf {\bibinfo {volume} {107}},\
  \bibinfo {pages} {053602} (\bibinfo {year} {2011})}\BibitemShut {NoStop}%
\bibitem [{\citenamefont {Liew}\ and\ \citenamefont
  {Savona}(2010)}]{PhysRevLett.104.183601}%
  \BibitemOpen
  \bibfield  {author} {\bibinfo {author} {\bibfnamefont {T.~C.~H.}\
  \bibnamefont {Liew}}\ and\ \bibinfo {author} {\bibfnamefont {V.}~\bibnamefont
  {Savona}},\ }\bibfield  {title} {\bibinfo {title} {Single photons from
  coupled quantum modes},\ }\href
  {https://doi.org/10.1103/PhysRevLett.104.183601} {\bibfield  {journal}
  {\bibinfo  {journal} {Phys. Rev. Lett.}\ }\textbf {\bibinfo {volume} {104}},\
  \bibinfo {pages} {183601} (\bibinfo {year} {2010})}\BibitemShut {NoStop}%
\bibitem [{\citenamefont {Bamba}\ \emph {et~al.}(2011)\citenamefont {Bamba},
  \citenamefont {Imamo\ifmmode~\breve{g}\else \u{g}\fi{}lu}, \citenamefont
  {Carusotto},\ and\ \citenamefont {Ciuti}}]{PhysRevA.83.021802}%
  \BibitemOpen
  \bibfield  {author} {\bibinfo {author} {\bibfnamefont {M.}~\bibnamefont
  {Bamba}}, \bibinfo {author} {\bibfnamefont {A.}~\bibnamefont
  {Imamo\ifmmode~\breve{g}\else \u{g}\fi{}lu}}, \bibinfo {author}
  {\bibfnamefont {I.}~\bibnamefont {Carusotto}},\ and\ \bibinfo {author}
  {\bibfnamefont {C.}~\bibnamefont {Ciuti}},\ }\bibfield  {title} {\bibinfo
  {title} {Origin of strong photon antibunching in weakly nonlinear photonic
  molecules},\ }\href {https://doi.org/10.1103/PhysRevA.83.021802} {\bibfield
  {journal} {\bibinfo  {journal} {Phys. Rev. A}\ }\textbf {\bibinfo {volume}
  {83}},\ \bibinfo {pages} {021802} (\bibinfo {year} {2011})}\BibitemShut
  {NoStop}%
\bibitem [{\citenamefont {Snijders}\ \emph {et~al.}(2018)\citenamefont
  {Snijders}, \citenamefont {Frey}, \citenamefont {Norman}, \citenamefont
  {Flayac}, \citenamefont {Savona}, \citenamefont {Gossard}, \citenamefont
  {Bowers}, \citenamefont {van Exter}, \citenamefont {Bouwmeester},\ and\
  \citenamefont {L\"offler}}]{Snijders2018PRL}%
  \BibitemOpen
  \bibfield  {author} {\bibinfo {author} {\bibfnamefont {H.~J.}\ \bibnamefont
  {Snijders}}, \bibinfo {author} {\bibfnamefont {J.~A.}\ \bibnamefont {Frey}},
  \bibinfo {author} {\bibfnamefont {J.}~\bibnamefont {Norman}}, \bibinfo
  {author} {\bibfnamefont {H.}~\bibnamefont {Flayac}}, \bibinfo {author}
  {\bibfnamefont {V.}~\bibnamefont {Savona}}, \bibinfo {author} {\bibfnamefont
  {A.~C.}\ \bibnamefont {Gossard}}, \bibinfo {author} {\bibfnamefont {J.~E.}\
  \bibnamefont {Bowers}}, \bibinfo {author} {\bibfnamefont {M.~P.}\
  \bibnamefont {van Exter}}, \bibinfo {author} {\bibfnamefont {D.}~\bibnamefont
  {Bouwmeester}},\ and\ \bibinfo {author} {\bibfnamefont {W.}~\bibnamefont
  {L\"offler}},\ }\bibfield  {title} {\bibinfo {title} {Observation of the
  unconventional photon blockade},\ }\href
  {https://doi.org/10.1103/PhysRevLett.121.043601} {\bibfield  {journal}
  {\bibinfo  {journal} {Phys. Rev. Lett.}\ }\textbf {\bibinfo {volume} {121}},\
  \bibinfo {pages} {043601} (\bibinfo {year} {2018})}\BibitemShut {NoStop}%
\bibitem [{\citenamefont {Vaneph}\ \emph {et~al.}(2018)\citenamefont {Vaneph},
  \citenamefont {Morvan}, \citenamefont {Aiello}, \citenamefont {F\'echant},
  \citenamefont {Aprili}, \citenamefont {Gabelli},\ and\ \citenamefont
  {Est\`eve}}]{Vaneph2018PRL}%
  \BibitemOpen
  \bibfield  {author} {\bibinfo {author} {\bibfnamefont {C.}~\bibnamefont
  {Vaneph}}, \bibinfo {author} {\bibfnamefont {A.}~\bibnamefont {Morvan}},
  \bibinfo {author} {\bibfnamefont {G.}~\bibnamefont {Aiello}}, \bibinfo
  {author} {\bibfnamefont {M.}~\bibnamefont {F\'echant}}, \bibinfo {author}
  {\bibfnamefont {M.}~\bibnamefont {Aprili}}, \bibinfo {author} {\bibfnamefont
  {J.}~\bibnamefont {Gabelli}},\ and\ \bibinfo {author} {\bibfnamefont
  {J.}~\bibnamefont {Est\`eve}},\ }\bibfield  {title} {\bibinfo {title}
  {Observation of the unconventional photon blockade in the microwave domain},\
  }\href {https://doi.org/10.1103/PhysRevLett.121.043602} {\bibfield  {journal}
  {\bibinfo  {journal} {Phys. Rev. Lett.}\ }\textbf {\bibinfo {volume} {121}},\
  \bibinfo {pages} {043602} (\bibinfo {year} {2018})}\BibitemShut {NoStop}%
\bibitem [{\citenamefont {Xu}\ and\ \citenamefont {Li}(2013)}]{Xu_2013}%
  \BibitemOpen
  \bibfield  {author} {\bibinfo {author} {\bibfnamefont {X.-W.}\ \bibnamefont
  {Xu}}\ and\ \bibinfo {author} {\bibfnamefont {Y.-J.}\ \bibnamefont {Li}},\
  }\bibfield  {title} {\bibinfo {title} {Antibunching photons in a cavity
  coupled to an optomechanical system},\ }\href
  {https://doi.org/10.1088/0953-4075/46/3/035502} {\bibfield  {journal}
  {\bibinfo  {journal} {J. Phys. B: At. Mol. Opt. Phys.}\ }\textbf {\bibinfo
  {volume} {46}},\ \bibinfo {pages} {035502} (\bibinfo {year}
  {2013})}\BibitemShut {NoStop}%
\bibitem [{\citenamefont {Savona}()}]{savona2013UPB}%
  \BibitemOpen
  \bibfield  {author} {\bibinfo {author} {\bibfnamefont {V.}~\bibnamefont
  {Savona}},\ }\href@noop {} {\bibinfo {title} {Unconventional photon blockade
  in coupled optomechanical systems}},\ \Eprint
  {https://arxiv.org/abs/1302.5937} {arXiv:1302.5937} \BibitemShut {NoStop}%
\bibitem [{\citenamefont {Zhang}\ \emph {et~al.}(2015)\citenamefont {Zhang},
  \citenamefont {Cheng}, \citenamefont {Liu},\ and\ \citenamefont
  {Zhou}}]{ZhangWZ2015PRA}%
  \BibitemOpen
  \bibfield  {author} {\bibinfo {author} {\bibfnamefont {W.-Z.}\ \bibnamefont
  {Zhang}}, \bibinfo {author} {\bibfnamefont {J.}~\bibnamefont {Cheng}},
  \bibinfo {author} {\bibfnamefont {J.-Y.}\ \bibnamefont {Liu}},\ and\ \bibinfo
  {author} {\bibfnamefont {L.}~\bibnamefont {Zhou}},\ }\bibfield  {title}
  {\bibinfo {title} {Controlling photon transport in the single-photon
  weak-coupling regime of cavity optomechanics},\ }\href
  {https://doi.org/10.1103/PhysRevA.91.063836} {\bibfield  {journal} {\bibinfo
  {journal} {Phys. Rev. A}\ }\textbf {\bibinfo {volume} {91}},\ \bibinfo
  {pages} {063836} (\bibinfo {year} {2015})}\BibitemShut {NoStop}%
\bibitem [{\citenamefont {Ferretti}\ \emph {et~al.}(2013)\citenamefont
  {Ferretti}, \citenamefont {Savona},\ and\ \citenamefont
  {Gerace}}]{Ferretti_2013}%
  \BibitemOpen
  \bibfield  {author} {\bibinfo {author} {\bibfnamefont {S.}~\bibnamefont
  {Ferretti}}, \bibinfo {author} {\bibfnamefont {V.}~\bibnamefont {Savona}},\
  and\ \bibinfo {author} {\bibfnamefont {D.}~\bibnamefont {Gerace}},\
  }\bibfield  {title} {\bibinfo {title} {Optimal antibunching in passive
  photonic devices based on coupled nonlinear resonators},\ }\href
  {https://doi.org/10.1088/1367-2630/15/2/025012} {\bibfield  {journal}
  {\bibinfo  {journal} {New J. Phys.}\ }\textbf {\bibinfo {volume} {15}},\
  \bibinfo {pages} {025012} (\bibinfo {year} {2013})}\BibitemShut {NoStop}%
\bibitem [{\citenamefont {Flayac}\ and\ \citenamefont
  {Savona}(2013)}]{PhysRevA.88.033836}%
  \BibitemOpen
  \bibfield  {author} {\bibinfo {author} {\bibfnamefont {H.}~\bibnamefont
  {Flayac}}\ and\ \bibinfo {author} {\bibfnamefont {V.}~\bibnamefont
  {Savona}},\ }\bibfield  {title} {\bibinfo {title} {Input-output theory of the
  unconventional photon blockade},\ }\href
  {https://doi.org/10.1103/PhysRevA.88.033836} {\bibfield  {journal} {\bibinfo
  {journal} {Phys. Rev. A}\ }\textbf {\bibinfo {volume} {88}},\ \bibinfo
  {pages} {033836} (\bibinfo {year} {2013})}\BibitemShut {NoStop}%
\bibitem [{\citenamefont {Gerace}\ and\ \citenamefont
  {Savona}(2014)}]{Gerace2014PRA}%
  \BibitemOpen
  \bibfield  {author} {\bibinfo {author} {\bibfnamefont {D.}~\bibnamefont
  {Gerace}}\ and\ \bibinfo {author} {\bibfnamefont {V.}~\bibnamefont
  {Savona}},\ }\bibfield  {title} {\bibinfo {title} {Unconventional photon
  blockade in doubly resonant microcavities with second-order nonlinearity},\
  }\href {https://doi.org/10.1103/PhysRevA.89.031803} {\bibfield  {journal}
  {\bibinfo  {journal} {Phys. Rev. A}\ }\textbf {\bibinfo {volume} {89}},\
  \bibinfo {pages} {031803} (\bibinfo {year} {2014})}\BibitemShut {NoStop}%
\bibitem [{\citenamefont {Xu}\ and\ \citenamefont
  {Li}(2014{\natexlab{a}})}]{PhysRevA.90.043822}%
  \BibitemOpen
  \bibfield  {author} {\bibinfo {author} {\bibfnamefont {X.-W.}\ \bibnamefont
  {Xu}}\ and\ \bibinfo {author} {\bibfnamefont {Y.}~\bibnamefont {Li}},\
  }\bibfield  {title} {\bibinfo {title} {Tunable photon statistics in weakly
  nonlinear photonic molecules},\ }\href
  {https://doi.org/10.1103/PhysRevA.90.043822} {\bibfield  {journal} {\bibinfo
  {journal} {Phys. Rev. A}\ }\textbf {\bibinfo {volume} {90}},\ \bibinfo
  {pages} {043822} (\bibinfo {year} {2014}{\natexlab{a}})}\BibitemShut
  {NoStop}%
\bibitem [{\citenamefont {Xu}\ and\ \citenamefont
  {Li}(2014{\natexlab{b}})}]{PhysRevA.90.033809}%
  \BibitemOpen
  \bibfield  {author} {\bibinfo {author} {\bibfnamefont {X.-W.}\ \bibnamefont
  {Xu}}\ and\ \bibinfo {author} {\bibfnamefont {Y.}~\bibnamefont {Li}},\
  }\bibfield  {title} {\bibinfo {title} {Strong photon antibunching of
  symmetric and antisymmetric modes in weakly nonlinear photonic molecules},\
  }\href {https://doi.org/10.1103/PhysRevA.90.033809} {\bibfield  {journal}
  {\bibinfo  {journal} {Phys. Rev. A}\ }\textbf {\bibinfo {volume} {90}},\
  \bibinfo {pages} {033809} (\bibinfo {year} {2014}{\natexlab{b}})}\BibitemShut
  {NoStop}%
\bibitem [{\citenamefont {Lemonde}\ \emph {et~al.}(2014)\citenamefont
  {Lemonde}, \citenamefont {Didier},\ and\ \citenamefont
  {Clerk}}]{Lemonde2014PRA}%
  \BibitemOpen
  \bibfield  {author} {\bibinfo {author} {\bibfnamefont {M.-A.}\ \bibnamefont
  {Lemonde}}, \bibinfo {author} {\bibfnamefont {N.}~\bibnamefont {Didier}},\
  and\ \bibinfo {author} {\bibfnamefont {A.~A.}\ \bibnamefont {Clerk}},\
  }\bibfield  {title} {\bibinfo {title} {Antibunching and unconventional photon
  blockade with gaussian squeezed states},\ }\href
  {https://doi.org/10.1103/PhysRevA.90.063824} {\bibfield  {journal} {\bibinfo
  {journal} {Phys. Rev. A}\ }\textbf {\bibinfo {volume} {90}},\ \bibinfo
  {pages} {063824} (\bibinfo {year} {2014})}\BibitemShut {NoStop}%
\bibitem [{\citenamefont {Shen}\ \emph {et~al.}(2015)\citenamefont {Shen},
  \citenamefont {Zhou},\ and\ \citenamefont {Yi}}]{PhysRevA.91.063808}%
  \BibitemOpen
  \bibfield  {author} {\bibinfo {author} {\bibfnamefont {H.~Z.}\ \bibnamefont
  {Shen}}, \bibinfo {author} {\bibfnamefont {Y.~H.}\ \bibnamefont {Zhou}},\
  and\ \bibinfo {author} {\bibfnamefont {X.~X.}\ \bibnamefont {Yi}},\
  }\bibfield  {title} {\bibinfo {title} {Tunable photon blockade in coupled
  semiconductor cavities},\ }\href {https://doi.org/10.1103/PhysRevA.91.063808}
  {\bibfield  {journal} {\bibinfo  {journal} {Phys. Rev. A}\ }\textbf {\bibinfo
  {volume} {91}},\ \bibinfo {pages} {063808} (\bibinfo {year}
  {2015})}\BibitemShut {NoStop}%
\bibitem [{\citenamefont {Zhou}\ \emph {et~al.}(2015)\citenamefont {Zhou},
  \citenamefont {Shen},\ and\ \citenamefont {Yi}}]{ZhouYH2015PRA}%
  \BibitemOpen
  \bibfield  {author} {\bibinfo {author} {\bibfnamefont {Y.~H.}\ \bibnamefont
  {Zhou}}, \bibinfo {author} {\bibfnamefont {H.~Z.}\ \bibnamefont {Shen}},\
  and\ \bibinfo {author} {\bibfnamefont {X.~X.}\ \bibnamefont {Yi}},\
  }\bibfield  {title} {\bibinfo {title} {Unconventional photon blockade with
  second-order nonlinearity},\ }\href
  {https://doi.org/10.1103/PhysRevA.92.023838} {\bibfield  {journal} {\bibinfo
  {journal} {Phys. Rev. A}\ }\textbf {\bibinfo {volume} {92}},\ \bibinfo
  {pages} {023838} (\bibinfo {year} {2015})}\BibitemShut {NoStop}%
\bibitem [{\citenamefont {Flayac}\ and\ \citenamefont
  {Savona}(2017)}]{PhysRevA.96.053810}%
  \BibitemOpen
  \bibfield  {author} {\bibinfo {author} {\bibfnamefont {H.}~\bibnamefont
  {Flayac}}\ and\ \bibinfo {author} {\bibfnamefont {V.}~\bibnamefont
  {Savona}},\ }\bibfield  {title} {\bibinfo {title} {Unconventional photon
  blockade},\ }\href {https://doi.org/10.1103/PhysRevA.96.053810} {\bibfield
  {journal} {\bibinfo  {journal} {Phys. Rev. A}\ }\textbf {\bibinfo {volume}
  {96}},\ \bibinfo {pages} {053810} (\bibinfo {year} {2017})}\BibitemShut
  {NoStop}%
\bibitem [{\citenamefont {Sarma}\ and\ \citenamefont
  {Sarma}(2017)}]{PhysRevA.96.053827}%
  \BibitemOpen
  \bibfield  {author} {\bibinfo {author} {\bibfnamefont {B.}~\bibnamefont
  {Sarma}}\ and\ \bibinfo {author} {\bibfnamefont {A.~K.}\ \bibnamefont
  {Sarma}},\ }\bibfield  {title} {\bibinfo {title}
  {Quantum-interference-assisted photon blockade in a cavity via parametric
  interactions},\ }\href {https://doi.org/10.1103/PhysRevA.96.053827}
  {\bibfield  {journal} {\bibinfo  {journal} {Phys. Rev. A}\ }\textbf {\bibinfo
  {volume} {96}},\ \bibinfo {pages} {053827} (\bibinfo {year}
  {2017})}\BibitemShut {NoStop}%
\bibitem [{\citenamefont {Zhou}\ \emph {et~al.}(2021)\citenamefont {Zhou},
  \citenamefont {Minganti}, \citenamefont {Qin}, \citenamefont {Wu},
  \citenamefont {Zhao}, \citenamefont {Fang}, \citenamefont {Nori},\ and\
  \citenamefont {Yang}}]{PhysRevA.104.053718}%
  \BibitemOpen
  \bibfield  {author} {\bibinfo {author} {\bibfnamefont {Y.~H.}\ \bibnamefont
  {Zhou}}, \bibinfo {author} {\bibfnamefont {F.}~\bibnamefont {Minganti}},
  \bibinfo {author} {\bibfnamefont {W.}~\bibnamefont {Qin}}, \bibinfo {author}
  {\bibfnamefont {Q.-C.}\ \bibnamefont {Wu}}, \bibinfo {author} {\bibfnamefont
  {J.-L.}\ \bibnamefont {Zhao}}, \bibinfo {author} {\bibfnamefont {Y.-L.}\
  \bibnamefont {Fang}}, \bibinfo {author} {\bibfnamefont {F.}~\bibnamefont
  {Nori}},\ and\ \bibinfo {author} {\bibfnamefont {C.-P.}\ \bibnamefont
  {Yang}},\ }\bibfield  {title} {\bibinfo {title} {$n$-photon blockade with an
  $n$-photon parametric drive},\ }\href
  {https://doi.org/10.1103/PhysRevA.104.053718} {\bibfield  {journal} {\bibinfo
   {journal} {Phys. Rev. A}\ }\textbf {\bibinfo {volume} {104}},\ \bibinfo
  {pages} {053718} (\bibinfo {year} {2021})}\BibitemShut {NoStop}%
\bibitem [{\citenamefont {Shen}\ \emph {et~al.}(2019)\citenamefont {Shen},
  \citenamefont {Qu}, \citenamefont {Li},\ and\ \citenamefont
  {Wu}}]{ShenS2019PRA}%
  \BibitemOpen
  \bibfield  {author} {\bibinfo {author} {\bibfnamefont {S.}~\bibnamefont
  {Shen}}, \bibinfo {author} {\bibfnamefont {Y.}~\bibnamefont {Qu}}, \bibinfo
  {author} {\bibfnamefont {J.}~\bibnamefont {Li}},\ and\ \bibinfo {author}
  {\bibfnamefont {Y.}~\bibnamefont {Wu}},\ }\bibfield  {title} {\bibinfo
  {title} {Tunable photon statistics in parametrically amplified photonic
  molecules},\ }\href {https://doi.org/10.1103/PhysRevA.100.023814} {\bibfield
  {journal} {\bibinfo  {journal} {Phys. Rev. A}\ }\textbf {\bibinfo {volume}
  {100}},\ \bibinfo {pages} {023814} (\bibinfo {year} {2019})}\BibitemShut
  {NoStop}%
\bibitem [{\citenamefont {Zubizarreta~Casalengua}\ \emph
  {et~al.}(2020)\citenamefont {Zubizarreta~Casalengua}, \citenamefont
  {López~Carreño}, \citenamefont {Laussy},\ and\ \citenamefont
  {Valle}}]{Zubizarreta2020LPR}%
  \BibitemOpen
  \bibfield  {author} {\bibinfo {author} {\bibfnamefont {E.}~\bibnamefont
  {Zubizarreta~Casalengua}}, \bibinfo {author} {\bibfnamefont {J.~C.}\
  \bibnamefont {López~Carreño}}, \bibinfo {author} {\bibfnamefont {F.~P.}\
  \bibnamefont {Laussy}},\ and\ \bibinfo {author} {\bibfnamefont {E.~d.}\
  \bibnamefont {Valle}},\ }\bibfield  {title} {\bibinfo {title} {Conventional
  and unconventional photon statistics},\ }\href
  {https://doi.org/https://doi.org/10.1002/lpor.201900279} {\bibfield
  {journal} {\bibinfo  {journal} {Laser Photon. Rev.}\ }\textbf {\bibinfo
  {volume} {14}},\ \bibinfo {pages} {1900279} (\bibinfo {year}
  {2020})}\BibitemShut {NoStop}%
\bibitem [{\citenamefont {Wang}\ \emph {et~al.}(2020)\citenamefont {Wang},
  \citenamefont {Bai}, \citenamefont {Liu}, \citenamefont {Zhang},\ and\
  \citenamefont {Wang}}]{Wang_2020NJP}%
  \BibitemOpen
  \bibfield  {author} {\bibinfo {author} {\bibfnamefont {D.-Y.}\ \bibnamefont
  {Wang}}, \bibinfo {author} {\bibfnamefont {C.-H.}\ \bibnamefont {Bai}},
  \bibinfo {author} {\bibfnamefont {S.}~\bibnamefont {Liu}}, \bibinfo {author}
  {\bibfnamefont {S.}~\bibnamefont {Zhang}},\ and\ \bibinfo {author}
  {\bibfnamefont {H.-F.}\ \bibnamefont {Wang}},\ }\bibfield  {title} {\bibinfo
  {title} {Photon blockade in a double-cavity optomechanical system with
  nonreciprocal coupling},\ }\href {https://doi.org/10.1088/1367-2630/abaa8a}
  {\bibfield  {journal} {\bibinfo  {journal} {New J. Phys.}\ }\textbf {\bibinfo
  {volume} {22}},\ \bibinfo {pages} {093006} (\bibinfo {year}
  {2020})}\BibitemShut {NoStop}%
\bibitem [{\citenamefont {Majumdar}\ \emph {et~al.}(2012)\citenamefont
  {Majumdar}, \citenamefont {Bajcsy}, \citenamefont {Rundquist},\ and\
  \citenamefont {Vu\ifmmode \check{c}\else
  \v{c}\fi{}kovi\ifmmode~\acute{c}\else \'{c}\fi{}}}]{PhysRevLett.108.183601}%
  \BibitemOpen
  \bibfield  {author} {\bibinfo {author} {\bibfnamefont {A.}~\bibnamefont
  {Majumdar}}, \bibinfo {author} {\bibfnamefont {M.}~\bibnamefont {Bajcsy}},
  \bibinfo {author} {\bibfnamefont {A.}~\bibnamefont {Rundquist}},\ and\
  \bibinfo {author} {\bibfnamefont {J.}~\bibnamefont {Vu\ifmmode \check{c}\else
  \v{c}\fi{}kovi\ifmmode~\acute{c}\else \'{c}\fi{}}},\ }\bibfield  {title}
  {\bibinfo {title} {Loss-enabled sub-poissonian light generation in a bimodal
  nanocavity},\ }\href {https://doi.org/10.1103/PhysRevLett.108.183601}
  {\bibfield  {journal} {\bibinfo  {journal} {Phys. Rev. Lett.}\ }\textbf
  {\bibinfo {volume} {108}},\ \bibinfo {pages} {183601} (\bibinfo {year}
  {2012})}\BibitemShut {NoStop}%
\bibitem [{\citenamefont {Zhang}\ \emph {et~al.}(2014)\citenamefont {Zhang},
  \citenamefont {Yu}, \citenamefont {Liu},\ and\ \citenamefont
  {Peng}}]{ZhangW2014PRA}%
  \BibitemOpen
  \bibfield  {author} {\bibinfo {author} {\bibfnamefont {W.}~\bibnamefont
  {Zhang}}, \bibinfo {author} {\bibfnamefont {Z.}~\bibnamefont {Yu}}, \bibinfo
  {author} {\bibfnamefont {Y.}~\bibnamefont {Liu}},\ and\ \bibinfo {author}
  {\bibfnamefont {Y.}~\bibnamefont {Peng}},\ }\bibfield  {title} {\bibinfo
  {title} {Optimal photon antibunching in a quantum-dot--bimodal-cavity
  system},\ }\href {https://doi.org/10.1103/PhysRevA.89.043832} {\bibfield
  {journal} {\bibinfo  {journal} {Phys. Rev. A}\ }\textbf {\bibinfo {volume}
  {89}},\ \bibinfo {pages} {043832} (\bibinfo {year} {2014})}\BibitemShut
  {NoStop}%
\bibitem [{\citenamefont {{Tang}}\ \emph {et~al.}(2015)\citenamefont {{Tang}},
  \citenamefont {{Geng}},\ and\ \citenamefont {{Xu}}}]{TangJ2015NatSR}%
  \BibitemOpen
  \bibfield  {author} {\bibinfo {author} {\bibfnamefont {J.}~\bibnamefont
  {{Tang}}}, \bibinfo {author} {\bibfnamefont {W.}~\bibnamefont {{Geng}}},\
  and\ \bibinfo {author} {\bibfnamefont {X.}~\bibnamefont {{Xu}}},\ }\bibfield
  {title} {\bibinfo {title} {{Quantum Interference Induced Photon Blockade in a
  Coupled Single Quantum Dot-Cavity System}},\ }\href
  {https://doi.org/10.1038/srep09252} {\bibfield  {journal} {\bibinfo
  {journal} {Sci. Rep.}\ }\textbf {\bibinfo {volume} {5}},\ \bibinfo {eid}
  {9252} (\bibinfo {year} {2015})}\BibitemShut {NoStop}%
\bibitem [{\citenamefont {Liang}\ \emph {et~al.}(2019)\citenamefont {Liang},
  \citenamefont {Duan}, \citenamefont {Guo}, \citenamefont {Liu}, \citenamefont
  {Guan},\ and\ \citenamefont {Ren}}]{LiangXY2019PRA}%
  \BibitemOpen
  \bibfield  {author} {\bibinfo {author} {\bibfnamefont {X.}~\bibnamefont
  {Liang}}, \bibinfo {author} {\bibfnamefont {Z.}~\bibnamefont {Duan}},
  \bibinfo {author} {\bibfnamefont {Q.}~\bibnamefont {Guo}}, \bibinfo {author}
  {\bibfnamefont {C.}~\bibnamefont {Liu}}, \bibinfo {author} {\bibfnamefont
  {S.}~\bibnamefont {Guan}},\ and\ \bibinfo {author} {\bibfnamefont
  {Y.}~\bibnamefont {Ren}},\ }\bibfield  {title} {\bibinfo {title}
  {Antibunching effect of photons in a two-level emitter-cavity system},\
  }\href {https://doi.org/10.1103/PhysRevA.100.063834} {\bibfield  {journal}
  {\bibinfo  {journal} {Phys. Rev. A}\ }\textbf {\bibinfo {volume} {100}},\
  \bibinfo {pages} {063834} (\bibinfo {year} {2019})}\BibitemShut {NoStop}%
\bibitem [{\citenamefont {Kyriienko}\ \emph {et~al.}(2020)\citenamefont
  {Kyriienko}, \citenamefont {Krizhanovskii},\ and\ \citenamefont
  {Shelykh}}]{PhysRevLett.125.197402}%
  \BibitemOpen
  \bibfield  {author} {\bibinfo {author} {\bibfnamefont {O.}~\bibnamefont
  {Kyriienko}}, \bibinfo {author} {\bibfnamefont {D.~N.}\ \bibnamefont
  {Krizhanovskii}},\ and\ \bibinfo {author} {\bibfnamefont {I.~A.}\
  \bibnamefont {Shelykh}},\ }\bibfield  {title} {\bibinfo {title} {{Nonlinear
  Quantum Optics with Trion Polaritons in 2D Monolayers: Conventional and
  Unconventional Photon Blockade}},\ }\href
  {https://doi.org/10.1103/PhysRevLett.125.197402} {\bibfield  {journal}
  {\bibinfo  {journal} {Phys. Rev. Lett.}\ }\textbf {\bibinfo {volume} {125}},\
  \bibinfo {pages} {197402} (\bibinfo {year} {2020})}\BibitemShut {NoStop}%
\bibitem [{\citenamefont {Wang}\ \emph {et~al.}(2021)\citenamefont {Wang},
  \citenamefont {Verstraelen}, \citenamefont {Zhang}, \citenamefont {Liew},\
  and\ \citenamefont {Chong}}]{WangY2021PRL}%
  \BibitemOpen
  \bibfield  {author} {\bibinfo {author} {\bibfnamefont {Y.}~\bibnamefont
  {Wang}}, \bibinfo {author} {\bibfnamefont {W.}~\bibnamefont {Verstraelen}},
  \bibinfo {author} {\bibfnamefont {B.}~\bibnamefont {Zhang}}, \bibinfo
  {author} {\bibfnamefont {T.~C.~H.}\ \bibnamefont {Liew}},\ and\ \bibinfo
  {author} {\bibfnamefont {Y.~D.}\ \bibnamefont {Chong}},\ }\bibfield  {title}
  {\bibinfo {title} {Giant enhancement of unconventional photon blockade in a
  dimer chain},\ }\href {https://doi.org/10.1103/PhysRevLett.127.240402}
  {\bibfield  {journal} {\bibinfo  {journal} {Phys. Rev. Lett.}\ }\textbf
  {\bibinfo {volume} {127}},\ \bibinfo {pages} {240402} (\bibinfo {year}
  {2021})}\BibitemShut {NoStop}%
\bibitem [{\citenamefont {Li}\ \emph {et~al.}(2024)\citenamefont {Li},
  \citenamefont {Hu},\ and\ \citenamefont {Yang}}]{li2023enhancement}%
  \BibitemOpen
  \bibfield  {author} {\bibinfo {author} {\bibfnamefont {J.}~\bibnamefont
  {Li}}, \bibinfo {author} {\bibfnamefont {C.-M.}\ \bibnamefont {Hu}},\ and\
  \bibinfo {author} {\bibfnamefont {Y.}~\bibnamefont {Yang}},\ }\bibfield
  {title} {\bibinfo {title} {Enhancement of photon blockade via topological
  edge states},\ }\href {https://doi.org/10.1103/PhysRevApplied.21.034058}
  {\bibfield  {journal} {\bibinfo  {journal} {Phys. Rev. Appl.}\ }\textbf
  {\bibinfo {volume} {21}},\ \bibinfo {pages} {034058} (\bibinfo {year}
  {2024})}\BibitemShut {NoStop}%
\bibitem [{\citenamefont {Lu}\ \emph {et~al.}()\citenamefont {Lu},
  \citenamefont {Wu},\ and\ \citenamefont {Lü}}]{lu2024chiral}%
  \BibitemOpen
  \bibfield  {author} {\bibinfo {author} {\bibfnamefont {Z.-G.}\ \bibnamefont
  {Lu}}, \bibinfo {author} {\bibfnamefont {Y.}~\bibnamefont {Wu}},\ and\
  \bibinfo {author} {\bibfnamefont {X.-Y.}\ \bibnamefont {Lü}},\ }\href@noop
  {} {\bibinfo {title} {Chiral interaction induced near-perfect photon
  blockade}},\ \Eprint {https://arxiv.org/abs/2402.09000} {arXiv:2402.09000}
  \BibitemShut {NoStop}%
\bibitem [{\citenamefont {Lingenfelter}\ \emph {et~al.}(2021)\citenamefont
  {Lingenfelter}, \citenamefont {Roberts},\ and\ \citenamefont
  {Clerk}}]{Andrew2021sciadv}%
  \BibitemOpen
  \bibfield  {author} {\bibinfo {author} {\bibfnamefont {A.}~\bibnamefont
  {Lingenfelter}}, \bibinfo {author} {\bibfnamefont {D.}~\bibnamefont
  {Roberts}},\ and\ \bibinfo {author} {\bibfnamefont {A.~A.}\ \bibnamefont
  {Clerk}},\ }\bibfield  {title} {\bibinfo {title} {Unconditional fock state
  generation using arbitrarily weak photonic nonlinearities},\ }\href
  {https://doi.org/10.1126/sciadv.abj1916} {\bibfield  {journal} {\bibinfo
  {journal} {Sci. Adv.}\ }\textbf {\bibinfo {volume} {7}},\ \bibinfo {pages}
  {eabj1916} (\bibinfo {year} {2021})}\BibitemShut {NoStop}%
\bibitem [{\citenamefont {Ma}\ and\ \citenamefont {Li}(2023)}]{MaYX2023PRA}%
  \BibitemOpen
  \bibfield  {author} {\bibinfo {author} {\bibfnamefont {Y.-X.}\ \bibnamefont
  {Ma}}\ and\ \bibinfo {author} {\bibfnamefont {P.-B.}\ \bibnamefont {Li}},\
  }\bibfield  {title} {\bibinfo {title} {Deterministic generation of phononic
  fock states via weak nonlinearities},\ }\href
  {https://doi.org/10.1103/PhysRevA.108.053709} {\bibfield  {journal} {\bibinfo
   {journal} {Phys. Rev. A}\ }\textbf {\bibinfo {volume} {108}},\ \bibinfo
  {pages} {053709} (\bibinfo {year} {2023})}\BibitemShut {NoStop}%
\bibitem [{\citenamefont {Zhou}\ \emph {et~al.}(2022)\citenamefont {Zhou},
  \citenamefont {Zhang}, \citenamefont {Liu}, \citenamefont {Wu}, \citenamefont
  {Shi}, \citenamefont {Shen},\ and\ \citenamefont {Yang}}]{ZhouYH2022PRAPP}%
  \BibitemOpen
  \bibfield  {author} {\bibinfo {author} {\bibfnamefont {Y.-H.}\ \bibnamefont
  {Zhou}}, \bibinfo {author} {\bibfnamefont {X.-Y.}\ \bibnamefont {Zhang}},
  \bibinfo {author} {\bibfnamefont {T.}~\bibnamefont {Liu}}, \bibinfo {author}
  {\bibfnamefont {Q.-C.}\ \bibnamefont {Wu}}, \bibinfo {author} {\bibfnamefont
  {Z.-C.}\ \bibnamefont {Shi}}, \bibinfo {author} {\bibfnamefont {H.-Z.}\
  \bibnamefont {Shen}},\ and\ \bibinfo {author} {\bibfnamefont {C.-P.}\
  \bibnamefont {Yang}},\ }\bibfield  {title} {\bibinfo {title} {Environmentally
  induced photon blockade via two-photon absorption},\ }\href
  {https://doi.org/10.1103/PhysRevApplied.18.064009} {\bibfield  {journal}
  {\bibinfo  {journal} {Phys. Rev. Appl.}\ }\textbf {\bibinfo {volume} {18}},\
  \bibinfo {pages} {064009} (\bibinfo {year} {2022})}\BibitemShut {NoStop}%
\bibitem [{\citenamefont {Su}\ \emph {et~al.}(2022)\citenamefont {Su},
  \citenamefont {Tang},\ and\ \citenamefont {Xia}}]{SuX2022PRA}%
  \BibitemOpen
  \bibfield  {author} {\bibinfo {author} {\bibfnamefont {X.}~\bibnamefont
  {Su}}, \bibinfo {author} {\bibfnamefont {J.-S.}\ \bibnamefont {Tang}},\ and\
  \bibinfo {author} {\bibfnamefont {K.}~\bibnamefont {Xia}},\ }\bibfield
  {title} {\bibinfo {title} {Nonlinear dissipation-induced photon blockade},\
  }\href {https://doi.org/10.1103/PhysRevA.106.063707} {\bibfield  {journal}
  {\bibinfo  {journal} {Phys. Rev. A}\ }\textbf {\bibinfo {volume} {106}},\
  \bibinfo {pages} {063707} (\bibinfo {year} {2022})}\BibitemShut {NoStop}%
\bibitem [{\citenamefont {Ben-Asher}\ \emph {et~al.}(2023)\citenamefont
  {Ben-Asher}, \citenamefont {Fern\'andez-Dom\'{\i}nguez},\ and\ \citenamefont
  {Feist}}]{Ben-Asher2023PRL}%
  \BibitemOpen
  \bibfield  {author} {\bibinfo {author} {\bibfnamefont {A.}~\bibnamefont
  {Ben-Asher}}, \bibinfo {author} {\bibfnamefont {A.~I.}\ \bibnamefont
  {Fern\'andez-Dom\'{\i}nguez}},\ and\ \bibinfo {author} {\bibfnamefont
  {J.}~\bibnamefont {Feist}},\ }\bibfield  {title} {\bibinfo {title}
  {Non-hermitian anharmonicity induces single-photon emission},\ }\href
  {https://doi.org/10.1103/PhysRevLett.130.243601} {\bibfield  {journal}
  {\bibinfo  {journal} {Phys. Rev. Lett.}\ }\textbf {\bibinfo {volume} {130}},\
  \bibinfo {pages} {243601} (\bibinfo {year} {2023})}\BibitemShut {NoStop}%
\bibitem [{\citenamefont {Zuo}\ \emph {et~al.}(2022)\citenamefont {Zuo},
  \citenamefont {Huang}, \citenamefont {Kuang}, \citenamefont {Xu},\ and\
  \citenamefont {Jing}}]{ZuoYL2022PRA}%
  \BibitemOpen
  \bibfield  {author} {\bibinfo {author} {\bibfnamefont {Y.}~\bibnamefont
  {Zuo}}, \bibinfo {author} {\bibfnamefont {R.}~\bibnamefont {Huang}}, \bibinfo
  {author} {\bibfnamefont {L.-M.}\ \bibnamefont {Kuang}}, \bibinfo {author}
  {\bibfnamefont {X.-W.}\ \bibnamefont {Xu}},\ and\ \bibinfo {author}
  {\bibfnamefont {H.}~\bibnamefont {Jing}},\ }\bibfield  {title} {\bibinfo
  {title} {Loss-induced suppression, revival, and switch of photon blockade},\
  }\href {https://doi.org/10.1103/PhysRevA.106.043715} {\bibfield  {journal}
  {\bibinfo  {journal} {Phys. Rev. A}\ }\textbf {\bibinfo {volume} {106}},\
  \bibinfo {pages} {043715} (\bibinfo {year} {2022})}\BibitemShut {NoStop}%
\bibitem [{\citenamefont {Kryuchkyan}\ \emph {et~al.}(2016)\citenamefont
  {Kryuchkyan}, \citenamefont {Shahinyan},\ and\ \citenamefont
  {Shelykh}}]{PhysRevA.93.043857}%
  \BibitemOpen
  \bibfield  {author} {\bibinfo {author} {\bibfnamefont {G.~Y.}\ \bibnamefont
  {Kryuchkyan}}, \bibinfo {author} {\bibfnamefont {A.~R.}\ \bibnamefont
  {Shahinyan}},\ and\ \bibinfo {author} {\bibfnamefont {I.~A.}\ \bibnamefont
  {Shelykh}},\ }\bibfield  {title} {\bibinfo {title} {Quantum statistics in a
  time-modulated exciton-photon system},\ }\href
  {https://doi.org/10.1103/PhysRevA.93.043857} {\bibfield  {journal} {\bibinfo
  {journal} {Phys. Rev. A}\ }\textbf {\bibinfo {volume} {93}},\ \bibinfo
  {pages} {043857} (\bibinfo {year} {2016})}\BibitemShut {NoStop}%
\bibitem [{\citenamefont {Hovsepyan}\ \emph {et~al.}(2014)\citenamefont
  {Hovsepyan}, \citenamefont {Shahinyan},\ and\ \citenamefont
  {Kryuchkyan}}]{PhysRevA.90.013839}%
  \BibitemOpen
  \bibfield  {author} {\bibinfo {author} {\bibfnamefont {G.~H.}\ \bibnamefont
  {Hovsepyan}}, \bibinfo {author} {\bibfnamefont {A.~R.}\ \bibnamefont
  {Shahinyan}},\ and\ \bibinfo {author} {\bibfnamefont {G.~Y.}\ \bibnamefont
  {Kryuchkyan}},\ }\bibfield  {title} {\bibinfo {title} {Multiphoton blockades
  in pulsed regimes beyond stationary limits},\ }\href
  {https://doi.org/10.1103/PhysRevA.90.013839} {\bibfield  {journal} {\bibinfo
  {journal} {Phys. Rev. A}\ }\textbf {\bibinfo {volume} {90}},\ \bibinfo
  {pages} {013839} (\bibinfo {year} {2014})}\BibitemShut {NoStop}%
\bibitem [{\citenamefont {Kyriienko}\ and\ \citenamefont
  {Liew}(2014)}]{PhysRevA.90.063805}%
  \BibitemOpen
  \bibfield  {author} {\bibinfo {author} {\bibfnamefont {O.}~\bibnamefont
  {Kyriienko}}\ and\ \bibinfo {author} {\bibfnamefont {T.~C.~H.}\ \bibnamefont
  {Liew}},\ }\bibfield  {title} {\bibinfo {title} {Triggered single-photon
  emitters based on stimulated parametric scattering in weakly nonlinear
  systems},\ }\href {https://doi.org/10.1103/PhysRevA.90.063805} {\bibfield
  {journal} {\bibinfo  {journal} {Phys. Rev. A}\ }\textbf {\bibinfo {volume}
  {90}},\ \bibinfo {pages} {063805} (\bibinfo {year} {2014})}\BibitemShut
  {NoStop}%
\bibitem [{\citenamefont {Ghosh}\ and\ \citenamefont
  {Liew}(2018)}]{PhysRevB.97.241301}%
  \BibitemOpen
  \bibfield  {author} {\bibinfo {author} {\bibfnamefont {S.}~\bibnamefont
  {Ghosh}}\ and\ \bibinfo {author} {\bibfnamefont {T.~C.~H.}\ \bibnamefont
  {Liew}},\ }\bibfield  {title} {\bibinfo {title} {Single photons from a gain
  medium below threshold},\ }\href {https://doi.org/10.1103/PhysRevB.97.241301}
  {\bibfield  {journal} {\bibinfo  {journal} {Phys. Rev. B}\ }\textbf {\bibinfo
  {volume} {97}},\ \bibinfo {pages} {241301} (\bibinfo {year}
  {2018})}\BibitemShut {NoStop}%
\bibitem [{\citenamefont {Ghosh}\ and\ \citenamefont
  {Liew}(2019)}]{PhysRevLett.123.013602}%
  \BibitemOpen
  \bibfield  {author} {\bibinfo {author} {\bibfnamefont {S.}~\bibnamefont
  {Ghosh}}\ and\ \bibinfo {author} {\bibfnamefont {T.~C.~H.}\ \bibnamefont
  {Liew}},\ }\bibfield  {title} {\bibinfo {title} {Dynamical blockade in a
  single-mode bosonic system},\ }\href
  {https://doi.org/10.1103/PhysRevLett.123.013602} {\bibfield  {journal}
  {\bibinfo  {journal} {Phys. Rev. Lett.}\ }\textbf {\bibinfo {volume} {123}},\
  \bibinfo {pages} {013602} (\bibinfo {year} {2019})}\BibitemShut {NoStop}%
\bibitem [{\citenamefont {Li}\ \emph {et~al.}(2022)\citenamefont {Li},
  \citenamefont {Zhang}, \citenamefont {Wu}, \citenamefont {Dong},
  \citenamefont {Zou}, \citenamefont {Guo},\ and\ \citenamefont
  {Zou}}]{PhysRevLett.129.043601}%
  \BibitemOpen
  \bibfield  {author} {\bibinfo {author} {\bibfnamefont {M.}~\bibnamefont
  {Li}}, \bibinfo {author} {\bibfnamefont {Y.-L.}\ \bibnamefont {Zhang}},
  \bibinfo {author} {\bibfnamefont {S.-H.}\ \bibnamefont {Wu}}, \bibinfo
  {author} {\bibfnamefont {C.-H.}\ \bibnamefont {Dong}}, \bibinfo {author}
  {\bibfnamefont {X.-B.}\ \bibnamefont {Zou}}, \bibinfo {author} {\bibfnamefont
  {G.-C.}\ \bibnamefont {Guo}},\ and\ \bibinfo {author} {\bibfnamefont {C.-L.}\
  \bibnamefont {Zou}},\ }\bibfield  {title} {\bibinfo {title} {Single-mode
  photon blockade enhanced by bi-tone drive},\ }\href
  {https://doi.org/10.1103/PhysRevLett.129.043601} {\bibfield  {journal}
  {\bibinfo  {journal} {Phys. Rev. Lett.}\ }\textbf {\bibinfo {volume} {129}},\
  \bibinfo {pages} {043601} (\bibinfo {year} {2022})}\BibitemShut {NoStop}%
\bibitem [{\citenamefont {Stefanatos}\ and\ \citenamefont
  {Paspalakis}(2020)}]{Stefanatos2020PRA}%
  \BibitemOpen
  \bibfield  {author} {\bibinfo {author} {\bibfnamefont {D.}~\bibnamefont
  {Stefanatos}}\ and\ \bibinfo {author} {\bibfnamefont {E.}~\bibnamefont
  {Paspalakis}},\ }\bibfield  {title} {\bibinfo {title} {{Dynamical blockade in
  a bosonic Josephson junction using optimal coupling}},\ }\href
  {https://doi.org/10.1103/PhysRevA.102.013716} {\bibfield  {journal} {\bibinfo
   {journal} {Phys. Rev. A}\ }\textbf {\bibinfo {volume} {102}},\ \bibinfo
  {pages} {013716} (\bibinfo {year} {2020})}\BibitemShut {NoStop}%
\bibitem [{\citenamefont {Kennedy}\ and\ \citenamefont
  {Eberhart}(1995)}]{488968}%
  \BibitemOpen
  \bibfield  {author} {\bibinfo {author} {\bibfnamefont {J.}~\bibnamefont
  {Kennedy}}\ and\ \bibinfo {author} {\bibfnamefont {R.}~\bibnamefont
  {Eberhart}},\ }\bibfield  {title} {\bibinfo {title} {Particle swarm
  optimization},\ }\href {https://doi.org/10.1109/ICNN.1995.488968} {\bibfield
  {journal} {\bibinfo  {journal} {IEEE Int. Conf. Neural Networks}\ }\textbf
  {\bibinfo {volume} {4}},\ \bibinfo {pages} {1942} (\bibinfo {year}
  {1995})}\BibitemShut {NoStop}%
\bibitem [{\citenamefont {Wang}\ \emph {et~al.}(2010)\citenamefont {Wang},
  \citenamefont {Lv}, \citenamefont {Zhu},\ and\ \citenamefont
  {Ma}}]{Wang2010PRB}%
  \BibitemOpen
  \bibfield  {author} {\bibinfo {author} {\bibfnamefont {Y.}~\bibnamefont
  {Wang}}, \bibinfo {author} {\bibfnamefont {J.}~\bibnamefont {Lv}}, \bibinfo
  {author} {\bibfnamefont {L.}~\bibnamefont {Zhu}},\ and\ \bibinfo {author}
  {\bibfnamefont {Y.}~\bibnamefont {Ma}},\ }\bibfield  {title} {\bibinfo
  {title} {Crystal structure prediction via particle-swarm optimization},\
  }\href {https://doi.org/10.1103/PhysRevB.82.094116} {\bibfield  {journal}
  {\bibinfo  {journal} {Phys. Rev. B}\ }\textbf {\bibinfo {volume} {82}},\
  \bibinfo {pages} {094116} (\bibinfo {year} {2010})}\BibitemShut {NoStop}%
\bibitem [{\citenamefont {Shokooh-Saremi}\ and\ \citenamefont
  {Magnusson}(2007)}]{Shokooh-Saremi:07}%
  \BibitemOpen
  \bibfield  {author} {\bibinfo {author} {\bibfnamefont {M.}~\bibnamefont
  {Shokooh-Saremi}}\ and\ \bibinfo {author} {\bibfnamefont {R.}~\bibnamefont
  {Magnusson}},\ }\bibfield  {title} {\bibinfo {title} {Particle swarm
  optimization and its application to the design of diffraction grating
  filters},\ }\href {https://doi.org/10.1364/OL.32.000894} {\bibfield
  {journal} {\bibinfo  {journal} {Opt. Lett.}\ }\textbf {\bibinfo {volume}
  {32}},\ \bibinfo {pages} {894} (\bibinfo {year} {2007})}\BibitemShut
  {NoStop}%
\bibitem [{\citenamefont {Zhang}\ and\ \citenamefont
  {Gong}(2019)}]{PhysRevB.100.235452}%
  \BibitemOpen
  \bibfield  {author} {\bibinfo {author} {\bibfnamefont {S.}~\bibnamefont
  {Zhang}}\ and\ \bibinfo {author} {\bibfnamefont {J.}~\bibnamefont {Gong}},\
  }\bibfield  {title} {\bibinfo {title} {Floquet engineering with particle
  swarm optimization: Maximizing topological invariants},\ }\href
  {https://doi.org/10.1103/PhysRevB.100.235452} {\bibfield  {journal} {\bibinfo
   {journal} {Phys. Rev. B}\ }\textbf {\bibinfo {volume} {100}},\ \bibinfo
  {pages} {235452} (\bibinfo {year} {2019})}\BibitemShut {NoStop}%
\bibitem [{\citenamefont {Stenberg}\ \emph {et~al.}(2016)\citenamefont
  {Stenberg}, \citenamefont {K\"ohn},\ and\ \citenamefont
  {Wilhelm}}]{PhysRevA.93.012122}%
  \BibitemOpen
  \bibfield  {author} {\bibinfo {author} {\bibfnamefont {M.~P.~V.}\
  \bibnamefont {Stenberg}}, \bibinfo {author} {\bibfnamefont {O.}~\bibnamefont
  {K\"ohn}},\ and\ \bibinfo {author} {\bibfnamefont {F.~K.}\ \bibnamefont
  {Wilhelm}},\ }\bibfield  {title} {\bibinfo {title} {Characterization of
  decohering quantum systems: Machine learning approach},\ }\href
  {https://doi.org/10.1103/PhysRevA.93.012122} {\bibfield  {journal} {\bibinfo
  {journal} {Phys. Rev. A}\ }\textbf {\bibinfo {volume} {93}},\ \bibinfo
  {pages} {012122} (\bibinfo {year} {2016})}\BibitemShut {NoStop}%
\bibitem [{\citenamefont {Prasad}\ and\ \citenamefont
  {Souradeep}(2012)}]{Prasad2012PRD}%
  \BibitemOpen
  \bibfield  {author} {\bibinfo {author} {\bibfnamefont {J.}~\bibnamefont
  {Prasad}}\ and\ \bibinfo {author} {\bibfnamefont {T.}~\bibnamefont
  {Souradeep}},\ }\bibfield  {title} {\bibinfo {title} {Cosmological parameter
  estimation using particle swarm optimization},\ }\href
  {https://doi.org/10.1103/PhysRevD.85.123008} {\bibfield  {journal} {\bibinfo
  {journal} {Phys. Rev. D}\ }\textbf {\bibinfo {volume} {85}},\ \bibinfo
  {pages} {123008} (\bibinfo {year} {2012})}\BibitemShut {NoStop}%
\bibitem [{\citenamefont {Liu}\ and\ \citenamefont {Liao}()}]{LiuYH2024arxiv}%
  \BibitemOpen
  \bibfield  {author} {\bibinfo {author} {\bibfnamefont {Y.-H.}\ \bibnamefont
  {Liu}}\ and\ \bibinfo {author} {\bibfnamefont {J.-Q.}\ \bibnamefont {Liao}},\
  }\href@noop {} {\bibinfo {title} {{Mechanical quadrature squeezing beyond the
  3dB limit via quantum learning control}}},\ \Eprint
  {https://arxiv.org/abs/2404.13563} {arXiv:2404.13563} \BibitemShut {NoStop}%
\bibitem [{\citenamefont {Carmichael}(1993)}]{Carmichael1993}%
  \BibitemOpen
  \bibfield  {author} {\bibinfo {author} {\bibfnamefont {H.}~\bibnamefont
  {Carmichael}},\ }\href {https://doi.org/10.1007/978-3-540-47620-7} {\emph
  {\bibinfo {title} {An Open Systems Approach to Quantum Optics}}}\ (\bibinfo
  {publisher} {Springer-Verlag Berlin Heidelberg},\ \bibinfo {year}
  {1993})\BibitemShut {NoStop}%
\bibitem [{\citenamefont {{Johansson}}\ \emph {et~al.}(2012)\citenamefont
  {{Johansson}}, \citenamefont {{Nation}},\ and\ \citenamefont
  {{Nori}}}]{QuTiP1}%
  \BibitemOpen
  \bibfield  {author} {\bibinfo {author} {\bibfnamefont {J.~R.}\ \bibnamefont
  {{Johansson}}}, \bibinfo {author} {\bibfnamefont {P.~D.}\ \bibnamefont
  {{Nation}}},\ and\ \bibinfo {author} {\bibfnamefont {F.}~\bibnamefont
  {{Nori}}},\ }\bibfield  {title} {\bibinfo {title} {{QuTiP: An open-source
  Python framework for the dynamics of open quantum systems}},\ }\href
  {https://doi.org/10.1016/j.cpc.2012.02.021} {\bibfield  {journal} {\bibinfo
  {journal} {Comp. Phys. Comm.}\ }\textbf {\bibinfo {volume} {183}},\ \bibinfo
  {pages} {1760} (\bibinfo {year} {2012})}\BibitemShut {NoStop}%
\bibitem [{\citenamefont {{Johansson}}\ \emph {et~al.}(2013)\citenamefont
  {{Johansson}}, \citenamefont {{Nation}},\ and\ \citenamefont
  {{Nori}}}]{QuTiP2}%
  \BibitemOpen
  \bibfield  {author} {\bibinfo {author} {\bibfnamefont {J.~R.}\ \bibnamefont
  {{Johansson}}}, \bibinfo {author} {\bibfnamefont {P.~D.}\ \bibnamefont
  {{Nation}}},\ and\ \bibinfo {author} {\bibfnamefont {F.}~\bibnamefont
  {{Nori}}},\ }\bibfield  {title} {\bibinfo {title} {{QuTiP 2: A Python
  framework for the dynamics of open quantum systems}},\ }\href
  {https://doi.org/10.1016/j.cpc.2012.11.019} {\bibfield  {journal} {\bibinfo
  {journal} {Comp. Phys. Comm.}\ }\textbf {\bibinfo {volume} {184}},\ \bibinfo
  {pages} {1234} (\bibinfo {year} {2013})}\BibitemShut {NoStop}%
\end{thebibliography}%
	
\end{document}